\documentclass[10pt,onecolumn,notitlepage,eqsecnum,floats,aps,amsmath,amssymb%
,nofootinbib,prd,showpacs]{revtex4-1}

\usepackage{amsmath,amssymb,amsfonts,amsthm}
\usepackage{enumerate}
\usepackage{graphicx}
\usepackage{psfrag}
\usepackage{subfigure}
\usepackage{calc}

\input cyracc.def
\newfam\cyrfam
\font\tencyr=wncyr10
\font\sevencyr=wncyr7
\font\fivecyr=wncyr5

\textfont\cyrfam=\tencyr \scriptfont\cyrfam=\sevencyr
\scriptscriptfont\cyrfam=\fivecyr

\newcommand{\ini}{{\rm in}}
\newcommand{\fin}{{\rm fin}}

\usepackage{LQC-symbols}

\begin{document}

\title{Inflation from non-minimally coupled scalar field in loop quantum cosmology}

\author{Micha{\l} \surname{Artymowski}${}^{1,2}$}
  \email{artymowski@fuw.edu.pl}

\author{Andrea \surname{Dapor}${}^1$}
  \email{adapor@fuw.edu.pl}
%   \affiliation{Instytut Fizyki Teoretycznej, Uniwersytet Warszawski,\\ 
%     ul. Ho\.{z}a 69, 00-681 Warszawa, Poland}

\author{Tomasz \surname{Paw{\l}owski}${}^{3,4}$}
  \email{tomasz.pawlowski@unab.cl}

\affiliation{${}^1$ Instytut Fizyki Teoretycznej, Uniwersytet Warszawski,
  ul. Ho\.{z}a 69, 00-681 Warszawa, Poland}
\affiliation{${}^2$ Department of Physics, Beijing Normal University,
  XinJieKouWai 19, 100875 Beijing, P.R. China.}
\affiliation{${}^3$ Departamento de Ciencias F\'isicas, Facultad de Ciencias Exactas,
  Universidad Andres Bello, Av.~Rep\'ublica 220,  Santiago de Chile.}
\affiliation{${}^4$ Katedra Metod Matematycznych Fizyki, Uniwersytet Warszawski,
    ul. Ho\.{z}a 74, 00-682 Warszawa, Poland}

\begin{abstract}
  The FRW model with non-minimally coupled {massive} scalar field has been investigated 
  in LQC framework. {Considered form of the potential and coupling allows applications 
  to Higgs driven inflation.} {Out of two frames used in the literature to describe 
  such systems: Jordan and Einstein frame, the latter one is applied. Specifically, 
  we explore the idea of the Einstein frame being the natural 'environment' for 
  quantization and the Jordan picture having an emergent nature.} The resulting dynamics
  qualitatively modifies the standard bounce paradigm in LQC in two ways: $(i)$ the 
  bounce point is no longer marked by critical matter energy density, $(ii)$ the Planck 
  scale physics features the ``mexican hat'' trajectory with two consecutive bounces 
  and rapid expansion and recollapse between them. {Furthermore, for physically viable 
  coupling strength and initial data the subsequent inflation exceeds $60$ e-foldings.}
\end{abstract}

% \keywords{{loop quantum cosmology, big bounce, inflation, Higgs-driven inflation}}
% \arxivnumber{\Red{1234.5678}}
% \pacs{}

\maketitle

\section{Introduction}

The inflation paradigm is one of the most successful ideas allowing to explain recent 
precise cosmological observations. However one of its main problems is the construction 
of the physically viable scenario featuring sufficiently long inflation epoch 
($>60$ e-foldings) with sufficiently large probability. In this context a lot of hope 
is attached to the models featuring a non-minimal coupling of gravity and a scalar field
(e.g. the Higgs field {\cite{Barvinsky:1998rn,*Qiu:2011tk,*Arai:2011nq,*Barvinsky:2009fy%
,Bezrukov:2010jz,Barvinsky:2008ia}}), 
vector fields \cite{Golovnev:2008cf} as well as fermions \cite{Channuie:2011rq} 
{(more precisely, the scalar degrees of freedom generated by the latter)}. {Such 
mechanism of driving the inflation is also efficient in generating the correct} 
primordial curvature perturbations. 

Among the considered scenarios the most popular one corresponds to a coupling of the 
form $\frac{1}{2}\xi\phi^2R$, where $R$ is the Ricci scalar {and $\phi$ is a scalar 
field}. The non-minimally coupled scalar inflaton drives the slow-roll inflation as 
long as the non-minimal{(ity of the)} coupling is strong, i.e. for $\xi\phi^2\gg1$. 
Values of self-coupling constant of scalar field {are determined by} the normalization 
of initial inhomogeneities. For example, {for the scalar field potential of the form} 
$V(\phi)={\frac{\lambda}{4}\left(\phi^2-2m^2/\lambda\right)^2}$ {the agreement with 
the observations (best fit) sets the self-coupling to} $\xi\sim47000\sqrt{\lambda}$. 
In this case, the mass term does not have a significant contribution {either} to the 
effective potential of the field, {or} to the normalization of inhomogeneities. Thus, 
$m$ may be of order of electroweak scale, as in the case of Higgs field {or a scalar 
originating from some extension of the Standard Model}. Thus, one is able to explain 
the whole cosmological data by a minimal amount of new physics. 
Note that besides some differences pointed out in \cite{Bezrukov:2011gp}, 
inflation from a non-minimally coupled scalar field is similar to inflation in 
$f(R)=R+\epsilon R^2$ gravity {\cite{Starobinsky:1980te}}.

{At this point it is worth noting that the type of the potential considered, 
while usually associated with the models of Higgs inflation, is not restricted 
just to this particular field. In fact, our studies can be easily (generalized and) 
applied to the analysis of the inflation driven by any non-minimally 
coupled scalar field with realistic values of $\xi$ and $\lambda$.} 

The analysis of the running of the coupling constant $\lambda$ limits the allowed 
range of $m$. In case of Higgs field $m\in\left(126 GeV,194 GeV\right)$ 
\cite{Bezrukov:2009db} with the theory error of order of 2 $GeV$. Thus, recent 
results from CMS \cite{CMS-PAS-HIG-12-020} and ATLAS \cite{ATLAS-CONF-2012-093} 
($m_H\simeq125$ GeV) are both consistent with SM Higgs inflation.

% The geometry 
While the considered model is very successful on the classical level it still 
suffers the standard problems related with the presence of initial singularity, 
which are expected to be solved by quantum gravity. One of the leading approaches to 
provide quantum description of spacetime itself is Loop Quantum Gravity (LQG) 
\cite{Rovelli-book,Thiemann-book,al-status}. The cosmological application of its 
symmetry reduced version, known as Loop Quantum Cosmology (LQC) \cite{bojo-liv,as-rev},
has indeed provided a qualitatively new picture of early Universe dynamics. The 
prediction of the so-called \emph{big bounce} phenomenon \cite{aps-prl} offered 
a new mechanism of resolving long standing cosmological problems. For example, the 
existence of a pre-bounce epoch of the Universe evolution provides an easy solution
to the horizon problem, while preliminary studies indicate that the 
dynamics in the near-bounce superinflation epoch prevents the catastrophic entropy
increase \cite{awe-entr,*bmp-gowdyeff-let,*bmp-gowdyeff-det} usually {considered a danger to}
bouncing cosmological models {(following the consideration of \cite{Tolman:1931zz})}. 
What is even more important, 
the spacetime discreteness effects amount to a dramatic increase of the probability
of inflation in the models with standard $m^2\phi^2$ potential scalar fields 
\cite{as-infl-let,*as-infl-det} (see also \cite{ck-infl}). Indeed for such models
the probability of inflation with enough $e$-foldings to ensure consistence with
$7$ years WMAP data happens with probability greater than $0.999997$. These results 
make the loop approach very attractive in inflationary cosmology.

The issue of LQC corrections to models non-minimally coupled to gravity has been 
analyzed in \cite{Bojowald:2006hd,*Bojowald:2006bz}, where the authors have introduced 
the LQC correction for small values of non-minimal coupling, {which were not 
consistent with observations}. The loop quantisation in the context of non-minimal 
coupling in scalar-tensor theories was considered {for example} in \cite{Zhang:2011vi}.
{In this paper we consider the model of FRW flat universe with scalar field 
coupled non-minimally to gravity with the realistic values of the constants $\lambda$, 
$\xi$. In consequence the model features all the regimes of: strong ($|\xi|\phi^2\gg1$), 
medium ($\xi^2\phi^2>1, \xi\phi^2<1$), and weak ($|\xi|\phi^2\ll1$) non-minimal 
coupling.\footnote{In this paper values of the inflaton are given in the units of 
  reduced Plack mass $M_{pl}\simeq 2.44\times 10^{18}GeV.$}
}

% Here already a comment about the methodology
In the literature the non-minimally coupled scalar field is usually investigated via 
two approaches: direct calculation in the frame defined by the action (called Jordan 
frame, see e.g.\cite{Steinwachs:2011zs}) and using certain conformal transformation 
to the frame in which the field is minimally coupled to gravity \cite{Bezrukov:2007ep}. 
It is claimed that these two approaches lead to nonequivalent results 
{\cite{Steinwachs:2011zs,Steinwachs:2013tr,Steinwachs-unpub}}. {The issue as to 
which one is the correct candidate for quantization is still open. In this paper we 
explore (with a slight modification) one of these possibilities, in order to test 
the viability\footnote{{Here as \emph{viable} we consider a model whose predictions 
  do not lead to immediate contradictions with observations.
}} of the model resulting from the choice. Namely, we probe} the idea, that the observed 
physics is native to Jordan frame \cite{Barvinsky:2008ia}, however we take the point 
of view that it is just an emergent theory, following from the underlying ``fundamental'' 
theory described by Einstein frame (see sect.~\ref{sec:Einstein} for more extensive 
discussion and justification). It is thus the latter one which is subject to the 
loop quantization in our work.

The structure of this paper {is the following:} In section \ref{sec:classical} we
{introduce the model and determine 
the resulting} equations of motion for a scalar field non-minimally coupled to
gravity. In section \ref{sec:Einstein} we perform a canonical transformation (to
the so-called Einstein frame), {which allows} us to recast the theory {as the 
one with} minimally coupled {matter}. In section \ref{sec:quantisation} the 
canonical quantization of {Ashtekar-Barbero} variables is carried out in the 
Einstein frame {via methods of LQC} {giving precisely defined quantum framework}. 
{This framework is next used in section \ref{sec:Effective} as the background for 
constructing the effective semiclassical description of the system's dynamics.
The resulting semiclassical equations of motion are then applied in 
section \ref{sec:methods} in the systematic analysis of the dynamical evolution 
of the universes described by the model.}
{Its results are presented in section \ref{sec:results}. Their general discussion 
involving in particular the treatment's limitations is provided in the concluding}
section \ref{sec:concl}.

\section{Non-minimally coupled scalar field}\label{sec:classical}

{We start with the general (non-symmetry reduced) system of gravity non-minimally
coupled to the massive scalar field, as defined by the following action}
\begin{equation}
  S[\phi,g_{\mu\nu}] 
  = \frac{1}{8 \pi G} \int d^4x\mathcal{L} 
  = \frac{1}{8 \pi G} \int d^{4}x\sqrt{-g}\left[-U(\phi)R+\frac{1}{2}(\partial_{\mu}\phi)(\partial^{\mu}\phi)-V(\phi)\right] \ , \label{eq:dzialanieogolne}
\end{equation}
where {the chosen metric tensor signature is  $(+,-,-,-)$,} 
$R$ is the Ricci scalar and $\phi$ is a scalar field (which can be the 
inflaton, Higgs field, modulus field etc). Notice that the limit of minimally 
coupled scalar field {corresponds to} $U(\phi)=1/2$. For the further studies 
we choose one of the most popular forms of coupling, namely 
\begin{equation}\label{eq:Udef}
  U(\phi)=\frac{1}{2}+\frac{1}{2}\xi\phi^2 .
\end{equation}
We will assume that such a non-minimally coupled scalar field $\phi$ 
{is the dominant matter content in} the Universe.

{We next reduce the above action to the case of flat FRW spacetime
$g = N^2\rd t^2 - a^2(t) {}^o\!q$} \footnote{The issue of a scalar 
  field non-minimally coupled to gravity with non-zero spatial 
  curvature was considered e.g. in \cite{Kamenshchik:1998bb}.
}. 
Then, after integration by parts the $Ua^2\ddot{a}$ term, one gets
{the reduced action}
\begin{equation}
  S = \frac{1}{8 \pi G} \int d^{4}x\ a^{3}\left[-6UH^{2}-6HU'\dot{\phi}
    +\frac{1}{2}\dot{\phi}^{2}-V\right] \ , \label{eq:dzialanie}
\end{equation}
{where $U'(\phi) := [\partial_{\phi}U](\phi)$.}

By {varying this} action with respect to $a$ and $\phi$, {we obtain the 
following} equations of motion:
\begin{subequations}\begin{align}
  \ddot{\phi}+3H\dot{\phi}+V' &= 6U'\left(\frac{\ddot{a}}{a}+H^{2}\right) \ , 
    \label{eq:ruchuphi}
  \\
  2U'\ddot{\phi}+2U\left(\frac{2\ddot{a}}{a}+H^{2}\right)
  +2U''\dot{\phi}^{2}+4HU'\dot{\phi}
  & = -8\pi G P, \label{eq:friedmanna2}
\end{align}\end{subequations}
where $P=\dot{\phi}^2/2-V$ is the pressure of the scalar field. {In the equations 
above the lapse function has been set to $N=1$, although at this level it can 
safely be left unfixed. The variation $\delta S/\delta N=0$ over it produces the
(symmetry-reduced)} scalar constraint
${\bf H}=0$, where ${\bf H}$ is the Hamiltonian of the scalar 
field, curvature and their coupling. To define it, let us introduce canonical 
momenta of the scalar field and of the scale factor:
\begin{equation}
  \pi_\phi = \frac{\partial\mathcal{L}}{\partial \dot{\phi}}=a^3\dot{\phi}-6U'\dot{a}a^2,
  \qquad
  \pi_a = \frac{\partial\mathcal{L}}{\partial \dot{a}}=-12U\dot{a}a-6U'\dot{\phi}a^2.\label{eq:pedykanoniczne}
\end{equation}
For $N=1$ the Hamiltonian {${\bf H}$ takes the form}
\begin{equation}
  {\bf H} = \pi_\phi\dot{\phi}+\pi_a\dot{a}-\mathcal{L}
  = -6U\dot{a}^2a-6U'\dot{\phi}\dot{a}a^2+a^3 8\pi G \rho,
  \label{eq:Hamiltonian}
\end{equation}
where $\rho$ is the energy density of the scalar field. 
{Then the constraint ${\bf H}=0$ implies}
\begin{equation}
  UH^{2}+HU'\dot{\phi}=\frac{8\pi G}{6}\rho \ . \label{eq:friedmanna1}
\end{equation}
{Finally, combining the} eq. (\ref{eq:ruchuphi}-\ref{eq:friedmanna1}) 
one obtains {the $2$nd order} equation of motion for $\phi$ {in which 
the curvature is felt only through the cosmic friction term}
\begin{equation}
  \ddot{\phi}+3H\dot{\phi}=\frac{2U'V-UV'-U'\dot{\phi}^{2}(3U''+\frac{1}{2})}{U+3U'^{2}}\ ,
  \label{eq:ruchuphidobre}
\end{equation} 
{where, as before ``${}'$'' denotes the derivative over $\phi$.}

\subsection*{Inflation from strong value of non-minimal coupling}

Let us {introduce the} slow-roll parameters {defined as follows}
\begin{equation}
  \epsilon = -\frac{\dot{H}}{H^2}\ ,
  \qquad
  \eta = \frac{\ddot{H}}{\dot{H}H}\ . 
  \label{eq:epsiloneta}
\end{equation}
{The universe is considered to be in the} slow-roll inflation epoch {whenever} 
$\epsilon,|\eta|\ll1$ {for an extended time}. {For the coupling as in
\eqref{eq:Udef} and the potential} $V(\phi)=\frac{\lambda}{4}\phi^4$ {these parameters 
can be estimated in the range of strong non-minimal coupling to be}
\begin{equation}
  \epsilon\propto\eta\propto\frac{1}{\xi\phi^2}\ . 
  \label{eq:slow-rollparameters}
\end{equation}
{It follows then} that one obtains slow-roll evolution {whenever} 
$\xi\phi^2>1$, so inflation ends {only} when the field leaves the 
strong coupling regime. 

{It is worth noting, that (in the considered scenario)}
during the inflation, $\phi\sim O(\xi^{-1/2})\ll M_{pl}$. This implies, that 
there is no need to consider non-renormalazible terms in the potential, 
{as here they}
shall be suppressed  by the Planck scale. This is a theoretical advantage of 
Higgs inflation over chaotic inflation models (in which $\phi>M_{pl}$ during 
inflation). {Also}, despite the Standard Model Higgs field being a complex 
one, the phase of a field may be expressed by a massless scalar degree of 
freedom (which does not have a significant influence on the evolution of 
space-time). It follows {then} that there is no qualitative difference between the 
dynamics of our simple model and of a more realistic Higgs-driven inflation. 
For details, see \cite{Kamenshchik:1998bb}. {In consequence, our considerations 
can be applied to the physically relevant case}

\section{Transformation to Einstein frame}\label{sec:Einstein}

{An alternative approach to describing the system defined in the previous section 
is its transformation to the so-called ``Einstein frame'', where, unlike in the 
original physical formulation (denoted in the literature as the ``Jordan frame''
\cite{Bezrukov:2010jz}), the matter is coupled minimally to gravity. This is 
achieved by the following (conformal) transformation of the metric tensor 
and the scalar field $(g_{\mu,\nu},\phi) \to (\tilde{g}_{\mu\nu},\tilde{\phi})$}
\begin{equation}
  \tilde{g}_{\mu\nu} = 2Ug_{\mu\nu}\ ,
  \qquad 
  \left(\frac{d\tilde{\phi}}{d\phi}\right)^2=\frac{1}{2}\frac{U+3U'^2}{U^2}\ .\label{eq:transformEin}
\end{equation}
{Under this change} $\tilde{S}\big[\tilde{g}_{\mu\nu},\tilde{\phi}\big] = 
S\left[g_{\mu\nu},\phi\right]$ and $\tilde{U}(\tilde{\phi})=1/2$, thus 
$\tilde{\phi}$ is {indeed} minimally coupled to gravity {(as required)}. 
The explicit relation between $\tilde{\phi}$ and $\phi$ may be found for all 
ranges of non-minimal coupling. The metric $\tilde{g}_{\mu\nu}$ can be 
{easily} expressed in the FRW form {in terms of} $\tilde{t}$ and 
$\tilde{a}$ defined by
\begin{equation}
  \tilde{a} = \sqrt{2U} a\ ,
  \qquad 
  {\rd\tilde{t} = \sqrt{2U} \rd t\ }.
  \label{eq:TildaTimeAndScale}
\end{equation}
Effective potential and energy density of $\tilde{\phi}$ {transform} as follows
\begin{equation}\label{eq:Vmin}
  \tilde{V}(\tilde{\phi}) = \frac{1}{4}\frac{V(\phi(\tilde{\phi}))}{U^2(\phi(\tilde{\phi}))}, \quad 
  \tilde{\rho} = \frac{1}{{4U}}\dot{\tilde{\phi}}^2+\tilde{V}\ .
\end{equation}
% where $N=\sqrt{2U}$ is a {(new)} lapse function \mnote{can't we simply write $\sqrt{2U}$ instead of N, in the eq above?}. 
The scalar constraint {has the standard form of GR minimally coupled to massive scalar 
field}, {thus one can directly apply the known procedure of canonical quantization in the LQC framework.} 
{Ashtekar-Barbero} variables in Einstein frame are of the form of
\begin{equation}
  \text{sgn}(\tilde{v}) \tilde{v} 
  = \frac{\tilde{a}^3}{2\pi \gamma \sqrt{\Delta} \ell_{Pl}^2}\ ,
  \qquad 
  \tilde{b} = -\gamma \sqrt{\Delta} \frac{1}{\tilde{a}} \frac{d\tilde{a}}{d\tilde{t}}\ ,\label{eq:pcEin}
\end{equation}
where $\Delta$ is the so-called \emph{area gap} \cite{aps-imp} which equals 
$\Delta=4\pi\gamma\sqrt{3}\lPl^2$ \cite{awe-entr} and $\gamma$ is the Barbero-Immirzi 
parameter \cite{Barbero:1994ap} equal to $\gamma = 0.2375\ldots$ \cite{dl-gamma,*m-gamma}. The 
Poisson bracket {between these variables is} $\{\tilde{v}, \tilde{b}\} = 2/\hbar$.\footnote{We use 
  variable $\tilde{b}$ with opposite sign than usual in order to regard $\tilde{v}$ as our 
  configuration variable.
}

% The variable choice motivation
Before proceeding with the quantization let us note that on the classical level 
we have at our disposal two sets of variables: $(v,b,\phi,\pi_\phi)$ corresponding 
to the Jordan frame and $(\tilde{v},\tilde{b},\tilde{\phi},\pi_{\tilde{\phi}})$ 
in the Einstein one.
Both sets are related with each other by canonical transformation, {thus classically 
at the level of specified models} the selection of either set is simply \emph{a matter 
of choice}. This may no longer be true, once the perturbations or perturbative 
quantum corrections are introduced \cite{Steinwachs:2011zs,Steinwachs-unpub}. 
On the other hand the discrete nature of loop quantization implies that its 
applications to both frames will most likely lead to \emph{inequivalent} results. 
The issue of which choice of the frame (in general) should be considered the correct, 
physical one is under investigation \cite{Steinwachs-unpub}, although at present 
it remains an open question. {This issue seems to induce a strong polarization 
within the cosmological community. Here we do not intend to take sides in this 
discussion. Instead, we choose one possibility and perform a preliminary study of 
the model resulting from this choice. Our goal is the initial viability test: 
verification whether within the minimalistic setting, and under further selection 
of the quantization method, the model does not lead to immediate contradictions 
with current cosmological observations. More precisely, we check whether and to 
what extent the model realizes the operative inflation paradigm {(which in turn features 
the closest agreement with the observations)}. {It often happens in physics that 
internal consistency of particular theory admits few choices, which next are 
discriminated on the physical level (via for example agreement with experiments).}
{Our choice here is the assumption} 
that the physical quantities are observed in Jordan frame.}

{Given the above choice, we introduce one further modification.}
Usually {the selection of} the set {(the frame)} corresponding to the physical, 
measured quantities {is the most natural}. In our case that would be the former set.
{Here however we would like to explore another (somewhat hybrid) approach: to consider 
the Jordan frame as the one corresponding to observed reality and in which all the 
measurable quantities should be calculated. However we will think about it as the 
``emergent'' formulation, which should arise} from the underlying theory 
{following from the quantization in} the Einstein frame. 
{The selection of an Einstein frame as the starting point for quantization has 
been applied (for other classes of nonminimally coupled fields) for example in 
\cite{rms-exit}.}

{Such approach stems from techniques employed for example in quantum field 
theory on curved spacetime where the fields are rescaled by a scale factor and 
the effective field subject to quantization mixes the physical degrees of freedom
of the geometry and matter (see for example \cite{Jacobson:2003vx}).}
{In our case it is further} motivated by two observations: one of practical, 
and the other of the philosophical (or rather aesthetic) nature. First, the 
Einstein frame is a much more natural candidate for quantization. Indeed, in 
that frame gravity and matter fields are minimally coupled and the gravitational 
part of the action takes the standard GR form {(Einstein-Hilbert action term)}. 
Thus, the quantization procedure is natural and well understood. Second, the 
classical structure (variables) underlying the loop quantization originates from 
classical GR and encodes its symmetries. Thus, provided that loop quantization is 
applicable to the considered system at all, Einstein frame is the natural arena 
for it. 

{One has to remember, however, that by themselves the above reasons
are not sufficient to select the approach as ``the right one''. As already stated, 
we consider it as one of few possibilities. Our goal is to test whether the 
choices made lead to a viable model worth further studies, or whether it shall 
be dismissed as unphysical.}
 
{The precise implementation of the idea is the following:}
\begin{enumerate}[(i)]
  \item First, we provide the complete quantum description in Einstein frame.
  \item Next, the quantum framework {is} used as the basis for constructing the 
    effective description, used in turn to analyze the dynamical behavior.
  \item Finally, the results of dynamics {are} translated to the 
    ``physical'' Jordan frame.
\end{enumerate}

\section{Loop quantization}\label{sec:quantisation}

As discussed above, we quantize the system in Einstein frame, choosing the polymer 
representation for the geometry degrees of freedom and the Schr\"odinger one for 
{the} matter {ones}.
Thus, the quantization procedure is a full analogy of the one applied to the FRW 
cosmology with minimally coupled massless scalar field in \cite{aps-imp}. 
% here explanation of 'hybrid' approach
{Such a ``hybrid'' approach can in principle be considered inconsistent, as 
geometry and matter are treated differently. Indeed, the strictly systematic 
procedure should involve the polymeric quantization of all the degrees of 
freedom. In full Loop Quantum Gravity the systematic polymeric quantization of 
the scalar field is known \cite{klb-scalar}. In context of the symmetry-reduced 
models it has also been studied {\cite{hk-scalar-uv,*kp-psc,ddl-scalar}}.
{In particular, the analysis of the massless scalar field in FRW universe performed 
in \cite{ddl-scalar} has shown that the dynamics of such system is \emph{identical}
to the one of the system where the scalar field has been quantized a la Schr\"odinger.}
{However, the polymer quantization of the scalar field} admits a critical 
difference in comparison to the geometry. The latter admits a precisely defined 
``loop scale'' -- essentially an energy scale at which quantum gravity effects become 
significant \cite{aps-imp}. Such precise scale follows directly from the fact that the first 
nonzero eigenvalue of the LQG area operator is well distinct from zero. In the 
case of the scalar field, the operator playing an analogous role has eigenvalues
arbitrarily close to zero \cite{klo-scalar}. As a consequence, the ``loop scale'' for 
the scalar field cannot be determined the way it is for the geometry. Furthermore, 
the spectral structure indicates that in such case taking the continuous limit 
within the loop quantization may be the correct approach, thus {effectively} leading to 
Schr\"odinger quantization. The issue is however still under study. For this reason, and 
due to the preliminary nature of our studies, we follow the ``hybrid approach'': 
polymeric quantum geometry and Schr\"odinger quantum matter. At this stage of 
the investigation such approach has the further advantage of the ability to directly 
compare our results with the existing LQC literature.}

{Let us now outline briefly the procedure itself.}
% kinematical quantization
To start with, we employ (the initial part of) the Dirac program, first quantizing 
the system on the kinematical level (ignoring the constraint). The resulting 
kinematical Hilbert space, $\Hil_{\kin}$, is of the form
\begin{equation}
  \Hil_{\kin} = \Hil_{\gr} \otimes \Hil_{\phi} , \quad
  \Hil_{\gr} = L^2(\bar{\re},\rd\mu_{\Bohr}) , \ \Hil_{\phi} = L^2(\re,\rd\phi) ,
\end{equation}
where $\bar{\re}$ is the Bohr compactification of the real line.
A convenient basis of $\Hil_{\gr}$ is formed by the eigenstates of the oriented 
volume operator $\hat{\tilde{V}} \equiv \widehat{\tilde{a}}^3$
\begin{equation}\label{eq:V-def}
  \hat{\tilde{V}}\ket{\tilde{v}} = 2\pi\gamma\sqrt{\Delta}\lPl^2 \hat{\tilde{v}} \ket{\tilde{v}} =: \alpha \ket{\tilde{v}}\ ,
\end{equation}
where $\ket{\tilde{v}}$ is eigenstate of operator $\hat{\tilde{v}}$ with eigenvalue $\tilde{v}$. They are orthonormal with respect to Kronecker delta, that is
$\ip{\tilde{v}}{\tilde{v}'} = \delta_{\tilde{v},\tilde{v}'}$.
On the other hand, the basis of $\Hil_{\phi}$ is provided by the generalized eigenstates 
$(\tilde{\phi}|$ of the operator $\hat{\tilde{\phi}}$ -- a quantum counterpart of 
the field value $\tilde{\phi}$:
\begin{equation}\label{eq:phi-def}
  \forall \chi\in\Hil_{\phi} \ (\tilde{\phi}|\hat{\tilde{\phi}} - \tilde{\phi} \hat{I}\ket{\chi} = 0  .
\end{equation}

% fiducial cell role
At this point a short comment regarding the exact physical meaning of the quantity 
$\tilde{V}$ is in order. In order to obtain a meaningful canonical description of an isotropic 
model with noncompact spatial slices (which is the case here), one introduces 
into the theory an infrared regulator -- some fiducial cell $\mathcal{V}$ constant in 
comoving coordinates. The quantity $\tilde{V}=\tilde{a}^3$ here corresponds exactly to the (oriented)
volume of that cell. Of course, one has to remember that in order to obtain consistent
description one has to make sure that the theory admits a well defined regulator{-}removal 
limit. 

% basic operators
Given $\Hil_{\kin}$, we now select the basic operators defined on some dense domain of it.
Two of them are the operators $\hat{\tilde{v}}$ and $\hat{\tilde{\phi}}$ (\ref{eq:V-def}, \ref{eq:phi-def}).
The remaining two are the unit shift operator $\hat{\tilde{N}}$ and the scalar field momentum
$\hat{\pi}_{\tilde{\phi}}$:
\begin{equation}
  \hat{\tilde{N}}\ket{\tilde{v}} = \ket{\tilde{v}+1} , \qquad  \hat{\pi}_{\tilde{\phi}} = i\hbar \partial_{\tilde{\phi}} .
\end{equation}
The pair $\hat{\tilde{v}}, \hat{\tilde{N}}$ is the equivalent of the operators of quantum flux and 
holonomy in full LQG, where the holonomy-flux algebra provides the basis for quantization 
\cite{Thiemann-book}. In particular, $\hat{\tilde{v}}$ represents (the power of) the flux across the unit surface
(in comoving coordinates), whereas $\hat{\tilde{N}}$ the holonomy along straight line \cite{abl-lqc}.

% quantum Hamiltonian constraint
The set $(\hat{\tilde{v}},\hat{\tilde{N}},\hat{\tilde{\phi}},\hat{\pi}_{\tilde{\phi}})$ is sufficient 
to construct the quantum counterpart of the Hamiltonian constraint -- the next step 
in Dirac program. The details of its construction are presented in \cite{aps-imp}. 
However, here we choose a bit different and more convenient symmetric factor ordering, 
the one used in \cite{hp-lqc}. The resulting operator is
\begin{subequations}\label{eq:Ham-quant}\begin{align}
  \hat{{\bf H}} &= \hat{{\bf H}}_{\gr}\otimes\id_{\phi} + \hat{{\bf H}}_{\phi} , &
  \hat{{\bf H}}_{\gr} &= \frac{3\pi G}{8\alpha} \sqrt{|\hat{\tilde{v}}|} (\hat{\tilde{N}}^2-\hat{\tilde{N}}^{-2})^2 \sqrt{|\hat{\tilde{v}}|}  ,
    \\ 
    &  &
  \hat{{\bf H}}_{\phi} &= \frac{1}{2\alpha}\widehat{|\tilde{v}|^{-1}}\pi_{\tilde{\phi}}^2 
    + \frac{\alpha}{\hbar} |\hat{\tilde{v}}| \tilde{V}(\tilde{\phi}) , 
\end{align}\end{subequations}
where $\alpha$ is defined via \eqref{eq:V-def} and $\tilde{V}(\tilde{\phi})$ is 
the effective potential given by \eqref{eq:Vmin}.

% completion of the Dirac program
Having at our disposal both $\Hil_{\kin}$ and the quantum constraint $\hat{{\bf H}}$ we 
can complete the Dirac program, constructing the physical Hilbert space (out of states 
annihilated by $\hat{{\bf H}}$) via group-averaging procedure \cite{almmt-gave}, and 
introducing the family of unitarily related observables to capture nontrivial 
information about the evolution of the described system. Such observables can be 
again constructed out of self-adjoint operators on $\Hil_{\kin}$ (kinematical 
observables) via group averaging. In generic situations, however, the form of the resulting
physical observables is very different {than} the original kinematical ones. Therefore,
to obtain a technically manageable theory, one usually resorts to the so-called
\emph{deparametrization}, introducing into the system a suitable matter field and using 
it as an internal clock. This is exactly the method originally used in \cite{aps-imp}.

% The dust deparametrization
One such deparametrization, particularly successful in full LQG, has been introduced 
in \cite{hp-lqg}. There, the time variable is provided by irrotational dust (see
\cite{bk-dust,*kt-dust} for studies of dust frames and \cite{gt-dust} for application 
of \emph{full} dust frame to LQG). Then, the synthesis of three elements (the specific 
matter field, the fixing of the time gauge provided by the
proper time of dust ``particles'', and the diffeomorphism-invariant formalism of 
standard LQG) allow to build a technically manageable completion of gravity 
quantization program. Its reduction to isotropic cosmology has been presented in 
\cite{hp-lqc}. Its main properties are: $(i)$ the Hamiltonian constraint becomes true Hamiltonian, $(ii)$ the kinematical Hilbert space $\Hil_{\kin}$ becomes 
(precisely) the physical one, so do the kinematical observables, and $(iii)$ the evolution
of the physical state is described by time-dependent Schr{\"o}dinger equation 
\begin{equation}
  -i\hbar\partial_{\tilde{t}} \Psi(\tilde{v},\tilde{\phi}) = \hat{{\bf H}} \Psi(\tilde{v},\tilde{\phi}) \ ,
\end{equation}
where the time $\tilde{t}$ provided by the gauge choice agrees with the cosmic time and 
the Hamiltonian $\hat{{\bf H}}$ (in our case given by \eqref{eq:Ham-quant}) is self-adjoint 
for any system of gravity coupled to non-exotic matter \cite{klp-ga}. This quantum description 
provides an excellent basis for further analysis of the dynamics. It can be performed 
on the genuine quantum level, although in this paper we will employ the so called 
\emph{effective description} which will be introduced in the next section.

% observables
In order to describe the dynamics we select the following set of observables, convenient 
in cosmology: scaled quantum volume $|\hat{\tilde{v}}|$, operator $\widehat{\sin(\tilde{b})}:=(i/2)[\hat{\tilde{N}}^2-\hat{\tilde{N}}^{-2}]$, field $\hat{\tilde{\phi}}$ and its momentum 
$\hat{\pi}_{\tilde{\phi}}$. All these operators, being originally kinematical observables in irrotational 
dust deparametrization automatically become physical.

Another observable especially relevant for cosmology and in particular for our further 
studies is an operator corresponding to Hubble parameter. It takes the form:
\begin{equation}\label{eq:Hubble-quant}
  \hat{\tilde{H}} = -\frac{1}{4i\gamma\sqrt{\Delta}}[\hat{\tilde{N}}^4-\hat{\tilde{N}}^{-4}] .
\end{equation}
 
% scalar time remark
At this point it is worth noting that the dust field, although convenient, is not 
the only possible choice of the internal time. One alternative, viable in the cosmological 
context is the choice of additional -- massless scalar field. Such field has been 
used exactly in the pioneering work \cite{aps-imp} and is in fact the most popular 
choice of clock in LQC. Its application to the full theory is well defined (as variation 
of the formalism of \cite{dgkl-gq}) at least formally, however the complications related 
with the properties of the scalar field {keep} its application to non-cosmological settings 
beyond current technical reach. In the homogeneous cosmology context it is manageable, 
although not free from (minor) technical difficulties, such as: 
$(i)$ evolution is provided by Klein-Gordon equation rather than Schr\"odinger one,
$(ii)$ in presence of any non-negative scalar field potential the Hamiltonian loses 
self-adjointness, and 
$(iii)$ the physical Hilbert space is a proper subspace of $\Hil_{\kin}$ and its 
identification requires explicit knowledge of spectral properties of $\hat{{\bf H}}$.
We discuss this formalism in more detail in appendix \ref{app:scalar-time}.

\section{Effective theory}\label{sec:Effective}

At this point we have at our disposal the physical Hilbert space, time, true Hamiltonian 
operator and the set of physical quantum observables. These components are sufficient 
to systematically study the dynamics of our system. The analysis on the 
genuine quantum level can be performed by the methods of \cite{aps-det}. However, since the aim of 
this paper is to obtain maximally complete picture of the semiclassical sector of 
the theory we resort to effective methods, leaving the verification of the results 
against the genuine quantum ones to future work. To arrive to such effective 
description, we employ the method introduced in \cite{bs-eff} (allowing in principle 
to account for arbitrary order quantum corrections, see e.g. \cite{bbhkm-eff}) in its $0$th order. 
In this order the resulting set of equations of motion is equivalent to the one 
provided by heuristic methods of the so called \emph{effective dynamics of LQC} 
\cite{sv-eff}. Its comparison against the genuine quantum dynamics in many LQC systems 
\cite{aps-imp,apsv-spher,*bp-negL,acs-slqc,ap-posL} has shown that it mimics the full 
quantum evolution to high level of accuracy. Furthermore, in some cases this result 
has been confirmed analytically \cite{t-eff}.

Technically, the method reduces to replacing the basic operators
in quantum Hamiltonian by their expectation values {evaluated on the semiclassical 
(sharply peaked) states\footnote{There we assume implicitly, that there exists 
a sufficiently large space of such states. Verification of this assumption requires 
testing the dynamics on the genuine quantum level.}}. In our case, this is
\begin{equation}
  \hat{\tilde{v}}^n \mapsto \expect{\hat{\tilde{v}}}^n , \qquad
  \hat{\tilde{N}} \mapsto \expect{\hat{\tilde{N}}} e^{i\tilde{b}/2} ,
\end{equation}
Applying this mapping to \eqref{eq:Ham-quant}, we get the following ``classical''
effective Hamiltonian
\begin{equation}\label{eq:Heff}
  {\bf H}_{\eff} = -\frac{3\pi G}{2\alpha}|\tilde{v}|\sin^2(\tilde{b}) + \frac{\pi_{\tilde{\phi}}^2}{2\alpha|\tilde{v}|} 
  + \frac{\alpha|\tilde{v}|}{\hbar} \tilde{V}(\tilde{\phi}) = E_{\eff}.
\end{equation}
where $E_{\eff}$ is the energy of the dust clock field.
The full set of the effective equations of motion is then provided by Hamilton's 
equations
\begin{subequations}\label{eq:eoms-Einst}\begin{align}
  \frac{d\tilde{v}}{d\tilde{t}} &= -\frac{6\pi G}{\alpha\hbar} |\tilde{v}| \sin(\tilde{b})\cos(\tilde{b}) &
  \frac{d\tilde{b}}{d\tilde{t}} &= \frac{3\pi G}{\alpha\hbar}\sin^2(\tilde{b}) + \frac{\pi_{\tilde{\phi}}^2}{\alpha\hbar \tilde{v}^2}
    - \frac{2\alpha}{\hbar^2} \tilde{V} \label{eq:eoms-geom} \\
  \frac{d\tilde{\phi}}{d\tilde{t}} &= \frac{\pi_{\tilde{\phi}}}{\alpha|\tilde{v}|} &
  \frac{d\pi_{\tilde{\phi}}}{d\tilde{t}} &= -\frac{\alpha|\tilde{v}|}{\hbar}\frac{\partial\tilde{V}}{\partial\tilde{\phi}} 
  \label{eq:eoms-matt}
\end{align}\end{subequations}
In order to eliminate the effects of {the dust field} on the dynamics, we set 
\begin{equation}\label{eq:M0}
  E_{\eff} = 0 ,
\end{equation}
thus reducing the presence of dust to a ``dust vacuum''. Under this condition the 
set of equations of motion \eqref{eq:eoms-geom} and (\ref{eq:Heff}, \ref{eq:M0})  
becomes equivalent to the following set
\begin{equation}
  3\tilde{H}^2 = 8\pi G \tilde{\rho}\left(1-\frac{\tilde{\rho}}{\rho_{cr}}\right)\ ,
  \qquad
  \frac{d\tilde{H}}{d\tilde{t}} 
  = -4\pi G \left(\tilde{\rho}+\tilde{P}\right)
    \left(1-2\frac{\tilde{\rho}}{\rho_{cr}}\right)\ ,\label{eq:friedmann1Ein}
\end{equation}
where $d\tilde{v}/d\tilde{t} = 3\tilde{v}\tilde{H}$ and $\rho_{cr} = 3/8\pi G \Delta \gamma^2$. The Big Bounce in Einstein frame appears for $\tilde{\rho}=\rho_{cr}$. 

It is worth noting that \eqref{eq:friedmann1Ein} can be converted to Jordan frame via \eqref{eq:transformEin}, {giving}
\begin{subequations}\begin{align}
  6UH^2+6U'H\dot{\phi}+\dot{\phi}^2\frac{3U'^2}{2U}
  &= 8\pi
  G\left(\rho+\dot{\phi}^2\frac{3U'^2}{2U}\right)\left(1-\frac{1}{4U^2\rho_{cr}}
    \left(\rho+\dot{\phi}^2\frac{3U'^2}{2U}\right)\right)\ ,
    \label{eq:friedmann1nonkanon}
  \\
  \dot{H}+H\frac{U'\dot{\phi}}{U}+\dot{\phi}^2\frac{2UU''-3U'^2}{4U^2}
  &= -8\pi G\dot{\phi}^2\frac{U+3U'^2}{4U^2}\left(1-\frac{1}{\rho_{cr}}
    \left(\dot{\phi}^2\frac{U+3U'^2}{4U^3}+\frac{V}{2U^2}\right)\right)\ .
    \label{eq:firedmann2nonkanon}
\end{align}\end{subequations}
Note that for $\xi \phi^2 \gg 1$, {unlike in the case of LQC with minimal coupling,} 
at the moment of the Bounce {we have} $\rho \gg \rho_{cr}$.

The equation of motion for $\phi$ is of the form (\ref{eq:ruchuphi}): surprisingly, 
the effective rolling force of the scalar field is not changed by the LQC correction. 
{and the} LQC correction does not influence the {classical} equation of 
motion for $\tilde{\phi}$, which (after the substitution of $\tilde{\phi}=\tilde{\phi}(\phi)$) 
gives eq. (\ref{eq:ruchuphidobre}).

\section{The analysis of the dynamics}\label{sec:methods}

% actual set of equations, numerical integration method
The set of equations of motion in the Einstein frame \eqref{eq:eoms-Einst} 
allows in principle to evaluate the time evolution of the canonical data 
$(\tilde{v},\tilde{b},\tilde{\phi},\pi_{\tilde{\phi}})$ up to the caveat that the 
form of $\phi(\tilde{\phi})$ is needed to specify $\tilde{V}(\tilde{\phi})$. 
This dependence is provided through the differential 
relation \eqref{eq:transformEin}. Therefore, for the sake of precision, it is much 
more convenient to formulate and integrate the mixed set of equations of motion 
for the variables $(\tilde{v},\tilde{b},\phi,\pi_{\tilde{\phi}})$, where the equation for 
$d\phi/dt$ follows from \eqref{eq:eoms-matt} and \eqref{eq:transformEin}. This set 
allows to determine all the relevant physical parameters. 
For technical reasons (faster evolution), it is more convenient to evolve 
the system with respect to the ``physical'' time $t$ of the Jordan frame. Using 
\eqref{eq:eoms-Einst}, we arrive to the final set of the evolution 
equations:
\begin{subequations}\label{eq:eoms-final}\begin{align}
  \frac{d\tilde{v}}{dt} &= -\frac{6\pi G}{\alpha\hbar} \sqrt{2U(\phi)} |\tilde{v}| \sin(\tilde{b})\cos(\tilde{b}) &
  \frac{d\tilde{b}}{dt} &= \sqrt{2U(\phi)} \left[ 
    \frac{3\pi G}{\alpha\hbar}\sin^2(\tilde{b}) + \frac{\pi_{\tilde{\phi}}^2}{\alpha\hbar \tilde{v}^2}
    - \frac{2\alpha}{\hbar^2} \tilde{V} 
  \right]\label{eq:eoms-geom-fin} \\
  \frac{d\phi}{dt} 
  &= \sqrt{\frac{4U^3(\phi)}{U(\phi)+3U'^2(\phi)}}  
  \frac{\pi_{\tilde{\phi}}}{\alpha|\tilde{v}|} &
  \frac{d\pi_{\tilde{\phi}}}{dt} &= -\frac{\alpha|\tilde{v}|}{\hbar} 
  \sqrt{\frac{4U^3(\phi)}{U(\phi)+3U'^2(\phi)}}
  \frac{\partial\tilde{V}}{\partial {\phi}}  .
  \label{eq:eoms-matt-fin}
\end{align}\end{subequations}

% initial conditions
In order to specify the initial conditions for the time evolution, we exploit the fact 
that each physical trajectory has a distinguished point, namely the big bounce in 
the Einstein frame (at which $\tilde{b}=\pi/2$). We chose it as our ``initial point'', 
setting $t=0$ there and evolving the initial data (set there) both forward and 
backward in time. Since the set \eqref{eq:eoms-final} is homogeneous in $\tilde{v}$, \footnote{One 
  needs to remember that $\tilde{v}$ encodes the information about the physical volume of 
  the chosen region of spacetime, whose choice for noncompact spatial topology of 
  the universe is arbitrary.
} we have the freedom in choosing its initial value. We set it to $1$. Having the 
geometry degrees of freedom set, we fix the scalar field momentum $\pi_{\tilde{\phi}}$
as the function $\pi_{\tilde{\phi}}(\tilde{v},\tilde{b},\phi,E_{\eff})$ using the Hamiltonian \eqref{eq:Heff}.  
To eliminate the influence of the dust ``clock'' field on the dynamics we further set its 
energy $E_{\eff}$ to zero. As a consequence we have at our disposal a $1$-parameter family 
of initial data $(t=0, \tilde{v}=1, \tilde{b}=\pi/2, \phi_{\ini},
\pi_{\tilde{\phi}}=\pi_{\tilde{\phi}}(\tilde{v}=1,\tilde{b}=\pi/2,\phi_{\ini},E_{\eff}=0))$ parametrized 
by the value $\phi_{\ini}$ of the scalar field $\phi$ at the bounce point in the 
Einstein frame.

% the integration
These initial data, supplied (and determined) by the set of constants $\xi,\lambda$, 
was then evolved using the fifth order adaptive Runge-Kutta (Cash-Carp) method using 
the effective dynamics module of the Numerical LQC library developed by T.~Paw{\l}owski
and J.~Olmedo. The raw results of the simulations were next postprocessed with the 
use of \texttt{Mathematica} and \texttt{gnuplot} software.

%  constants and range of simulations
In actual simulations the value of $\lambda$ was set to $1/2$, which is the same order of magnitude as in the Standard Model. Due to the running of the coupling constant one cannot predict the precise value of $\lambda$ around the Planck scale. However recent results from ATLAS and CMS \cite{CMS-PAS-HIG-12-020,ATLAS-CONF-2012-093} suggest that $\lambda$ shall not be too big, since the mass of the Higgs is close to the minimal allowed value which does not violate electroweak vacuum. Thus, the value of $\lambda$ assumed by us is realistic.

The relation between coupling constants $\xi$ and $\lambda$ is given by the normalization of primordial inhomogeneities, which gives $\xi=47000\sqrt{\lambda}$. However, to confirm 
the robustness of the results and further analyze the qualitative behavior of the 
system, several different values have been considered: the sequence of lower values 
$\xi_n=\{4.7\times 10^n \sqrt{\lambda}; n=0,...,3\}$.
To test the wide range of trajectories the initial scalar field values were chosen 
from the interval $\phi_{\ini} \in (-10\xi^{-1/2} , 10\xi^{-1/2})$. For most 
simulations we selected $20$ initial points distributed unformly within this interval.
These initial data have been next evolved till the time $t_{\fin}=2\times 10^3,...,10^5$ 
depending on the simulation (and, in particular, on the constant $\xi$).
The results of these simulations are discussed in the next section.

\section{The results}\label{sec:results}

\begin{figure}[tbh!]
  \psfrag{phi}{$\phi$}
  \psfrag{H}{$H$}
  \psfrag{N}{$N$}
  \subfigure[]{
    \includegraphics[scale=0.60]{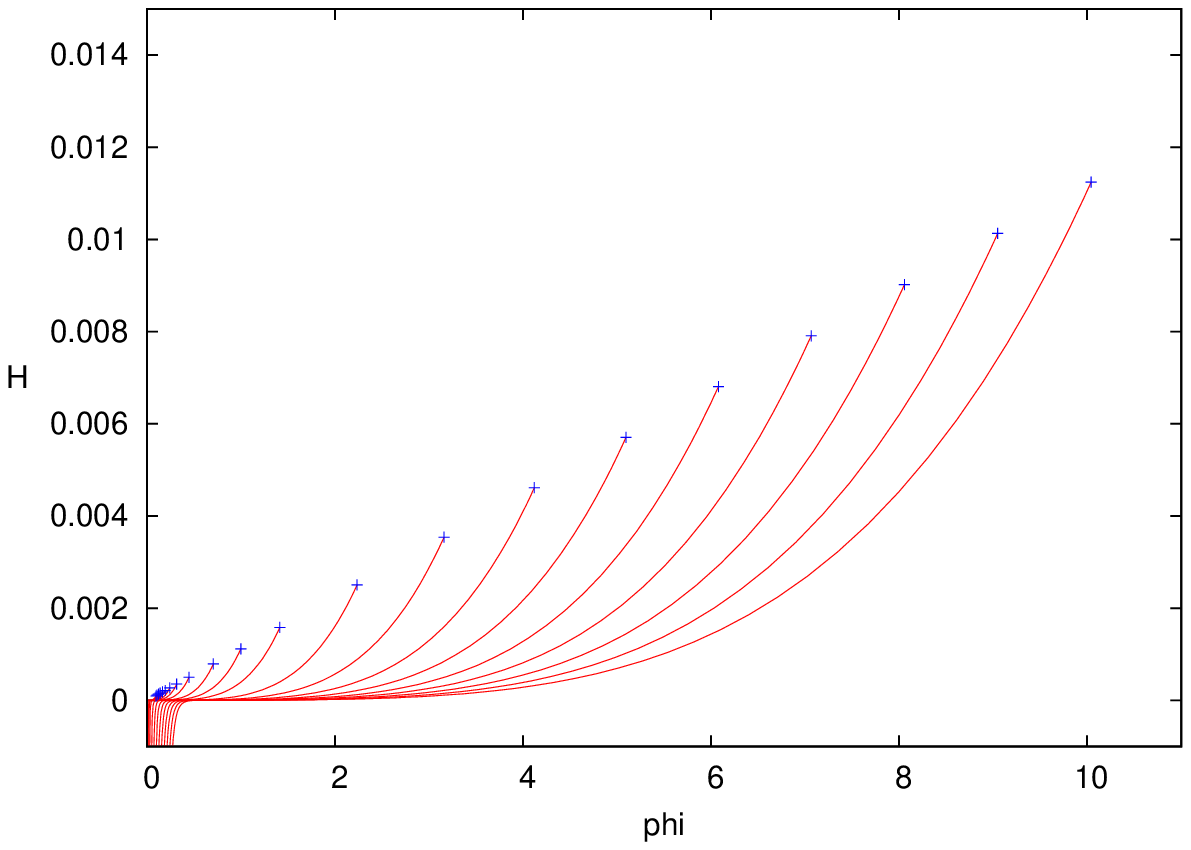}
    \label{fig:phase47K-hub-zoom3}
  }
  \subfigure[]{
    \includegraphics[scale=0.9]{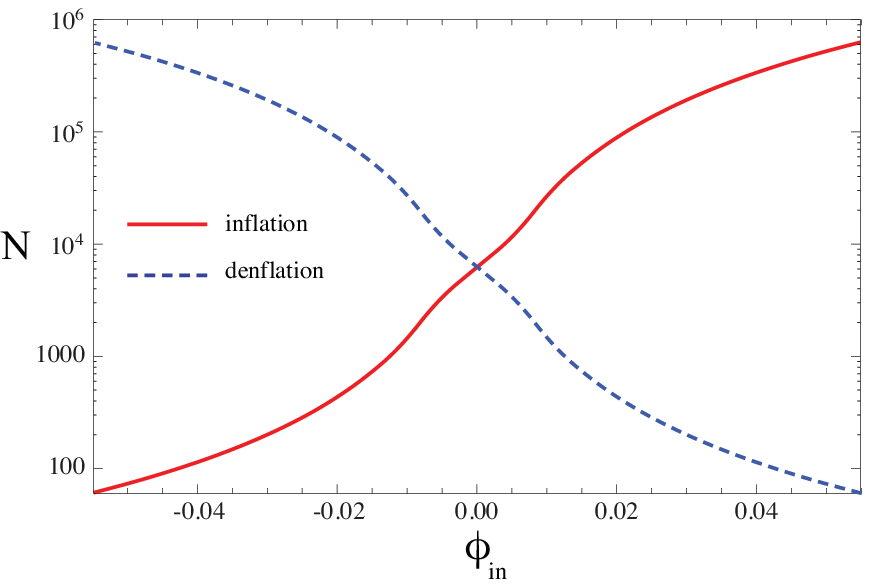}
    \label{fig:e-folds}
  }
  \caption{\ref{fig:phase47K-hub-zoom3} Projection of the phase portrait onto the 
    $\phi$--$H$ plane. The trajectories reach the onset of inflation (marked by 
    blue crosses) which points lie on a straight line (as expected in the strong 
    non-minimal coupling case, where $H\simeq\pm\sqrt{\lambda/12\xi}|\phi|$). 
    \ref{fig:e-folds} shows the number of $e$-foldings during the slow-roll inflation 
    after the bounces (red line) and the deflation before the bounces (blue line) 
    as a function of initial 
    value of the field (labeling the trajectories). Notice that even the trajectories 
    with very high time reversal asymmetry (considered to be unlikely) 
    give more than $60$ e-foldings during inflation.
  }
  \label{fig:phase47K-hub-zooms}
\end{figure}

\begin{figure}[tbh!]
  \psfrag{t}{$t$}
  \psfrag{H}{$H$}
  \subfigure[]{
    \includegraphics[scale=0.9]{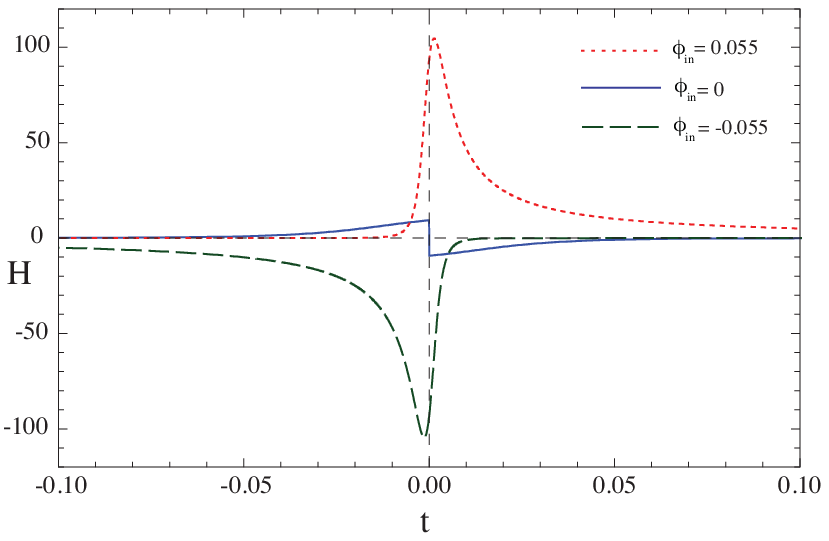}
    \label{fig:xi=47k-Hubble1}
  }
  \subfigure[]{
    \includegraphics[scale=0.60]{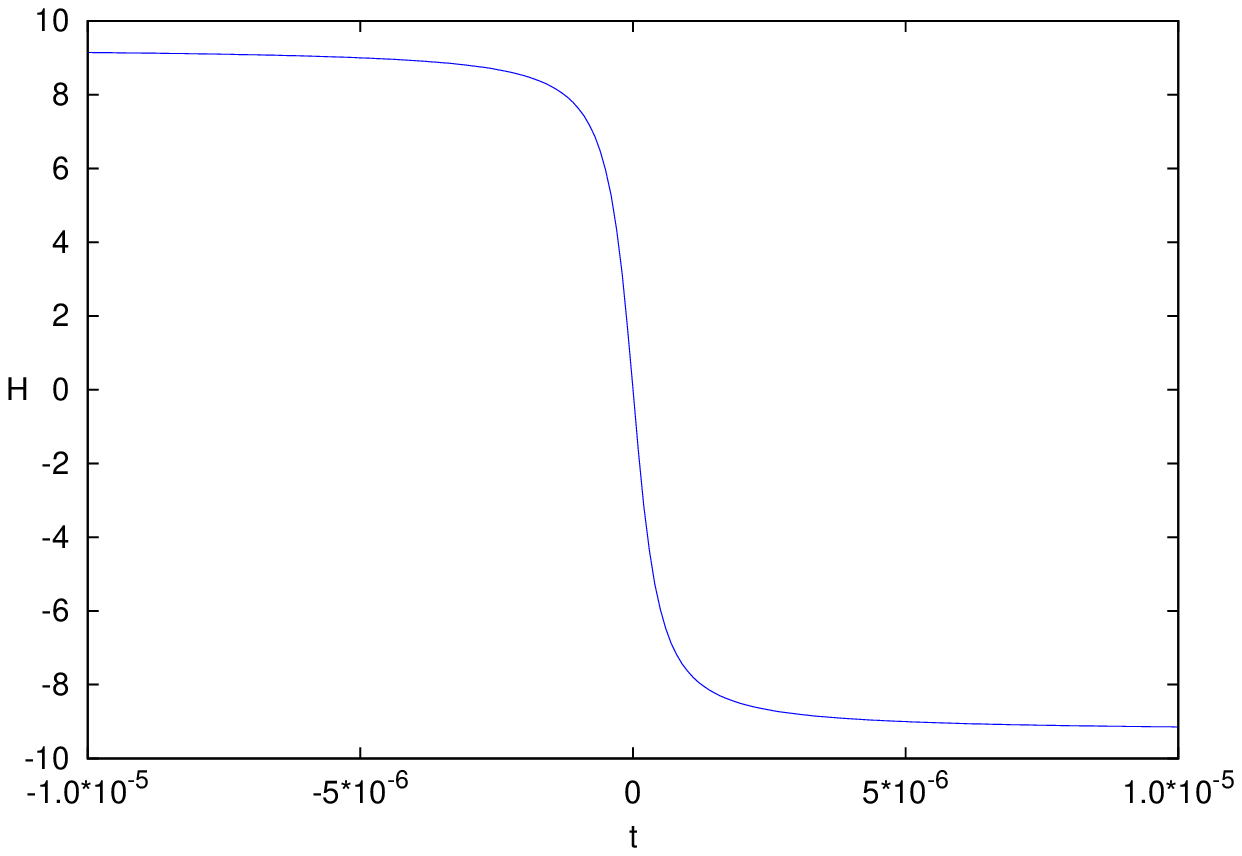}
    \label{fig:Htraj47K-10-zoom1}
  }
  \subfigure[]{
    \includegraphics[scale=0.9]{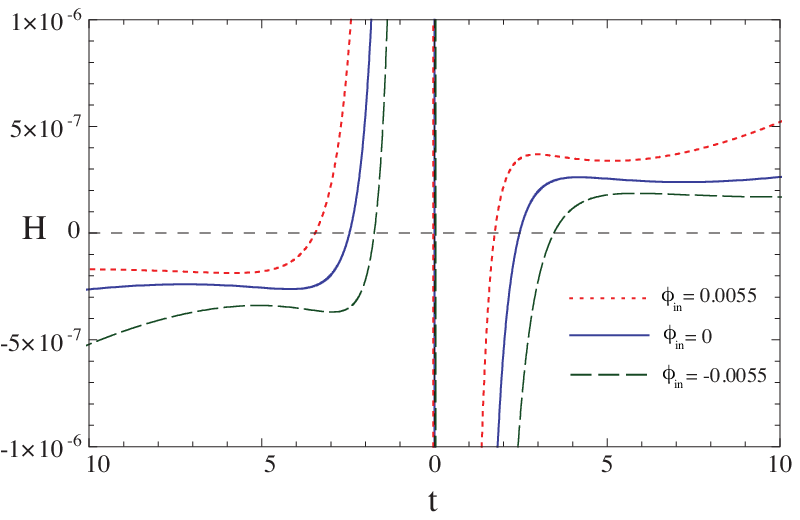}
    \label{fig:xi=47k-Hubble2}
  }
  \subfigure[]{
    \includegraphics[scale=0.9]{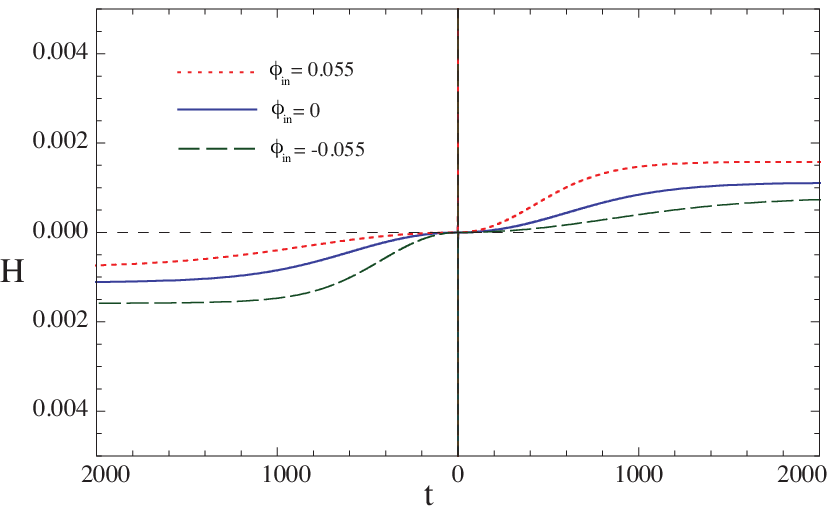}
    \label{fig:xi=47k-Hubble3}
  }
  \caption{Evolution of the Hubble parameter as a function of time (in Planck
    units) for the symmetric trajectory ($\phi_{\text{in}} = 0$) and for two non-symmetric
    ones. 
    Fig.~\ref{fig:xi=47k-Hubble1} shows the general view around the epoch of high 
    energy density.
    In \ref{fig:Htraj47K-10-zoom1} we zoom around $t = 0$ for the symmetric
    trajectory only (with the recollapse at $t=0$), to show that the transition 
    is in fact smooth.
    \ref{fig:xi=47k-Hubble2} visualizes two bounces in Jordan frame. 
    \ref{fig:xi=47k-Hubble3} shows the regions of the onset of inflation and the end
    of deflation, where $H$ is approximately constant.
  }
  \label{fig:Htraj47K-10}
\end{figure}

\begin{figure}[tbh!]
  \subfigure[]{
    \includegraphics[scale=0.9]{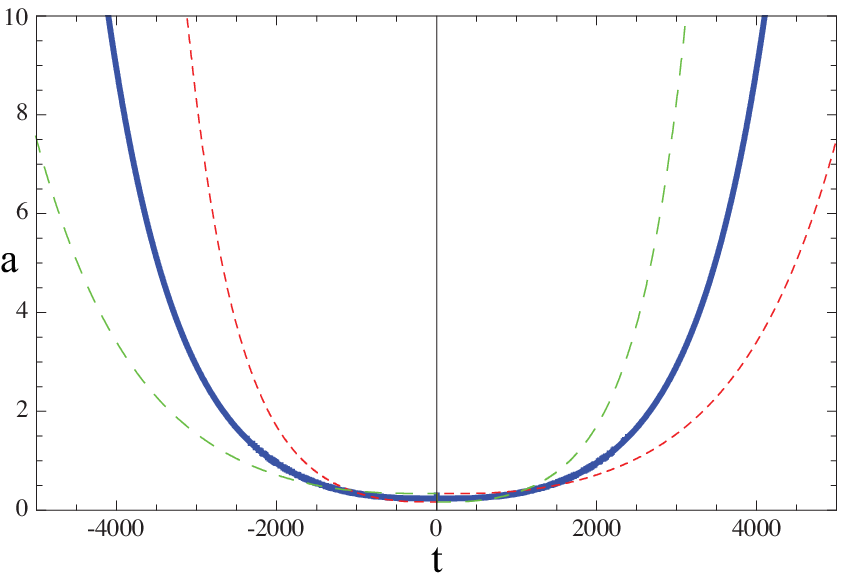}
    \label{fig:xi=47k-Scale1}
  }
  \subfigure[]{
    \includegraphics[scale=0.9]{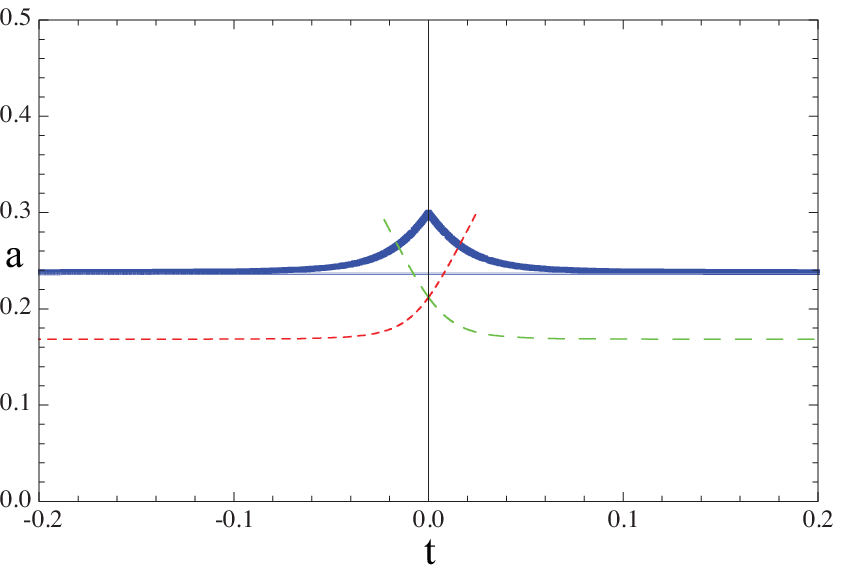}
    \label{fig:xi=47k-Scale2}
  }
  \caption{The evolution of the scale factor for the symmetric trajectory and for
    two non-symmetric ones ($\phi_{\text{in}} = \pm 0.0055$) \ref{fig:xi=47k-Scale1} 
    and its close-up around $t = 0$ \ref{fig:xi=47k-Scale2} showing the characteristic
    ``mexican hat'' shape with two bounces and recollapse between them. Note that
    for non-symmetric trajectories the ``tip of the hat'' (i.e., the moment of
    recollapse) is shifted in time and less evident.}
  \label{fig:phase47K-scale-zooms}
\end{figure}

\begin{figure}[tbh!]
  \subfigure[]{
    \includegraphics[scale=0.9]{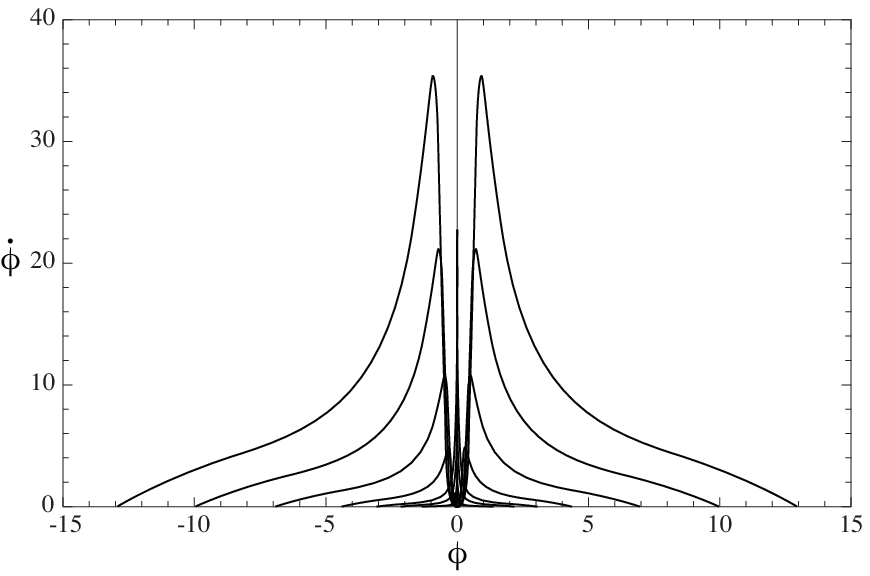}
    \label{fig:xi=47-PhaseSpace1}
  }
  \subfigure[]{
    \includegraphics[scale=0.9]{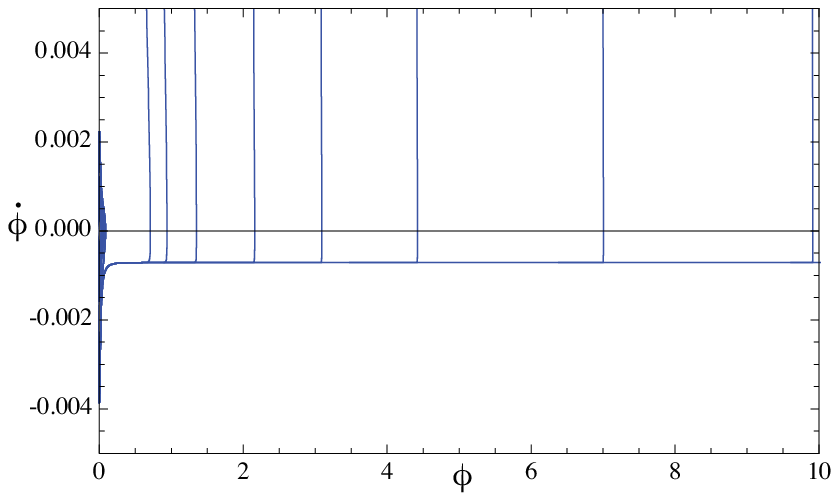}
    \label{fig:xi=47-PhaseSpace3.InflationAttractor}
  }
  \subfigure[]{
    \includegraphics[scale=0.9]{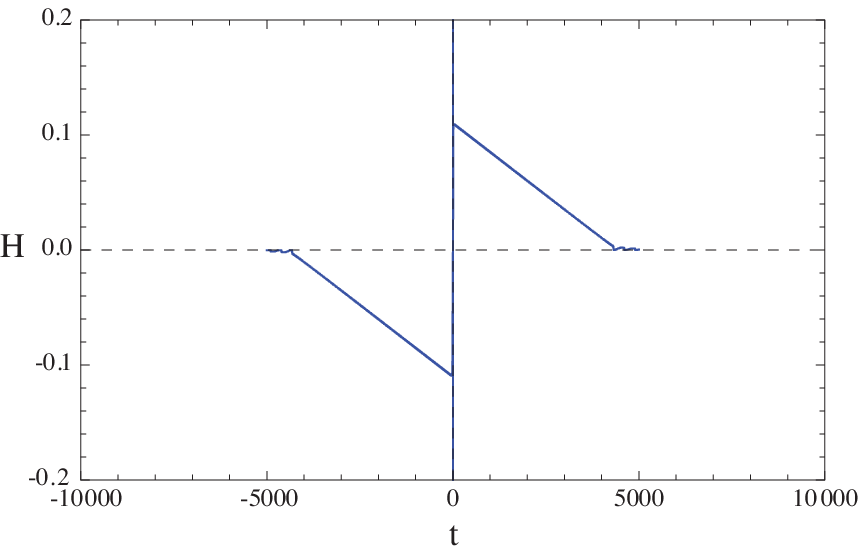}
    \label{fig:xi=47-HubbleParameterSymmetricTrajectory1}
  }
  \subfigure[]{
    \includegraphics[scale=0.9]{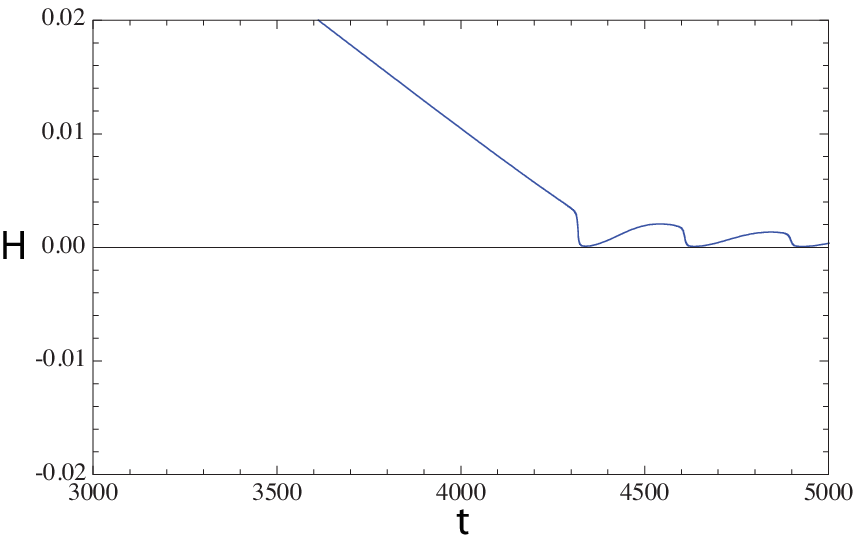}
    \label{xi=47-HubbleParameterSymmetricTrajectory6}
  }
  \caption{Phase space (matter sector: $\phi$--$\dot{\phi}$ plane) and evolution 
    of Hubble parameter for the $\xi = 47 \sqrt{\lambda}$ toy model. For such a 
    small value of the coupling we are able to evolve the system through the 
    whole inflation.
    Fig.~\ref{fig:xi=47-PhaseSpace1} shows the global view of the $\dot{\phi}$ half.
    \ref{fig:xi=47-PhaseSpace3.InflationAttractor} shows the approach of the trajectories 
    towards the attractor (ending in the reheating at $\dot{\phi} \sim \sqrt{\lambda}\phi^2$). 
    In \ref{fig:xi=47-HubbleParameterSymmetricTrajectory1} we present the time evolution 
    of the Hubble parameter during the inflation/deflation epoch. 
    \ref{xi=47-HubbleParameterSymmetricTrajectory6} is a close-up of the same evolution 
    around reheating. The qualitative features demonstrated here are expected to be also 
    present in the realistic $\xi = 47000 \sqrt{\lambda}$ case.
  }
  \label{fig:phase47-globalandzooms}
\end{figure}

The results of our studies can be summarized in the following set of points:
\begin{itemize}
  \item In the (underlying) Einstein frame we observe the standard for LQC picture 
    of a single bounce separating contracting and expanding epochs of the universe 
    evolution. The ``bare'' matter energy density $\tilde{\rho}$ and 
    Hubble parameter are bounded by their respective critical values 
    $\rho_c \approx 0.41\rho_{\rm Pl}$ and {$H_c \approx 1.3\lPl^{-1}$}.
  \item In the Jordan frame, here conjectured to represent the observed dynamics, 
    the standard bounce paradigm is slightly {changed}. The single large energy 
    density epoch still separates two long epochs of contraction and expansion. 
    The process of the bounce itself is however modified: the evolution of the 
    scale factor features the so called ``mexican hat'' shape (see 
    Fig.~\ref{fig:phase47K-scale-zooms}) -- the sequence of
    bounce, ultrarapid nonadiabatic expansion ending with recollapse, similar 
    nonadiabatic epoch of contraction and the final (second) bounce, after which 
    the universe expands to the classical regime (see also Fig.~\ref{fig:Htraj47K-10} 
    for the Hubble parameter evolution). The time between the bounces is of the order
    of Planck time.
  \item Outside of the ``mexican hat bounce'' the dynamical trajectory approaches 
    quickly the one predicted by GR. In particular in the future of the bounce 
    the value of the field $\phi$ grows to certain (depending on the trajectory) 
    maximal value $\phi_{\rm infl}$ at which point the slow roll inflation starts.
    Due to the time symmetry of the equations of motion the inflation after the 
    bounce is accompanied by the ``slow roll'' deflation before it. The initial
    Hubble parameter at the onset of inflation is proportional to $\phi_{\rm infl}$ 
    (see Fig.~\ref{fig:phase47K-hub-zoom3}).
  \item The number of e-foldings during the inflation is estimated to be 
    $N_{+}\simeq 3/16 \xi\phi_{\rm infl}^2$ and for the physical value 
    of $\xi\approx 4.7\cdot 10^4$
    exceeds $60$ within all the studied range of initial data (see Fig.~\ref{fig:e-folds}). 
    The same estimate holds for the deflation: $N_{-}\simeq 3/16 \xi\phi_{\rm defl}^2$, 
    where $\phi_{\rm defl}$ is the minimal value of the field (reached before the bounce).
  \item Within the precision of our estimates, the product $N_{+}N_{-}$ of the 
    e-foldings during inflation and deflation does not depend on the initial data. 
    It is the function of $\xi$ and $\lambda$ only, in particular growing with $\xi$. 
    For the selected values of $\xi,\lambda$ it equals $N_{+}N_{-} = {\left[4.4\cdot 10^3\right]^2}$
    (see Fig.~\ref{fig:e-folds}). This also implies that to have an inflation with $N < 60$, 
    there must be a deflation with $N > 10^6$. This is an extremely asymmetric situation, 
    and therefore highly unlikely {at least on the intuitive level: we thus expect
    that the} inflation with $N > 60$ is extremely probable - {{with much higher
    probability than} in} the {case of the} standard chaotic inflation.
  \item At the late time each trajectory reaches the standard inflationary attractor 
    and the reheating (see Fig.~\ref{fig:phase47-globalandzooms} for the illustration 
    on the example of the unphysical model corresponding to $\xi=47 \sqrt{\lambda}$).
\end{itemize}

\begin{figure}[tbh!]
  \psfrag{phi}{$\phi$}
  \psfrag{dphi}{$\dot{\phi}$}
  \subfigure[]{
    \includegraphics[scale=0.60]{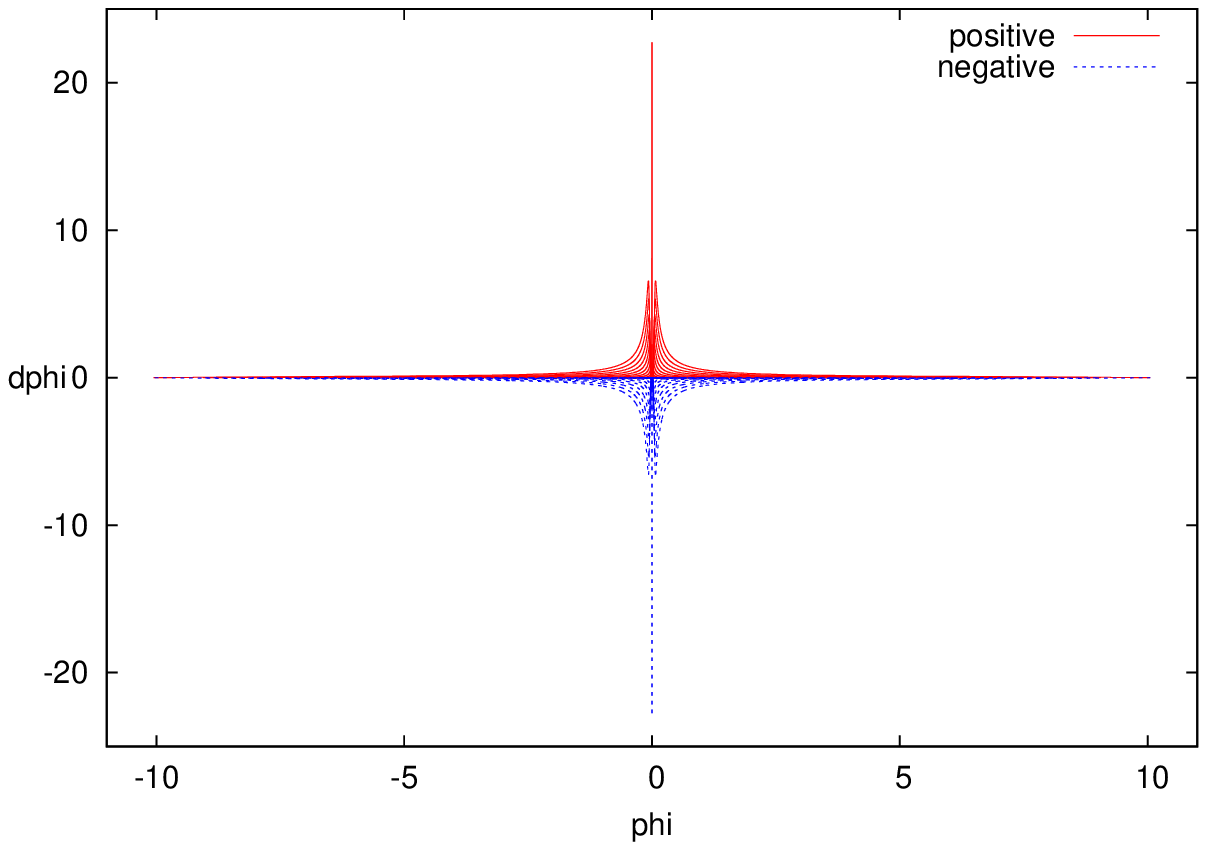}
    \label{fig:phase47K-global-matt}
  }
  \subfigure[]{
    \includegraphics[scale=0.60]{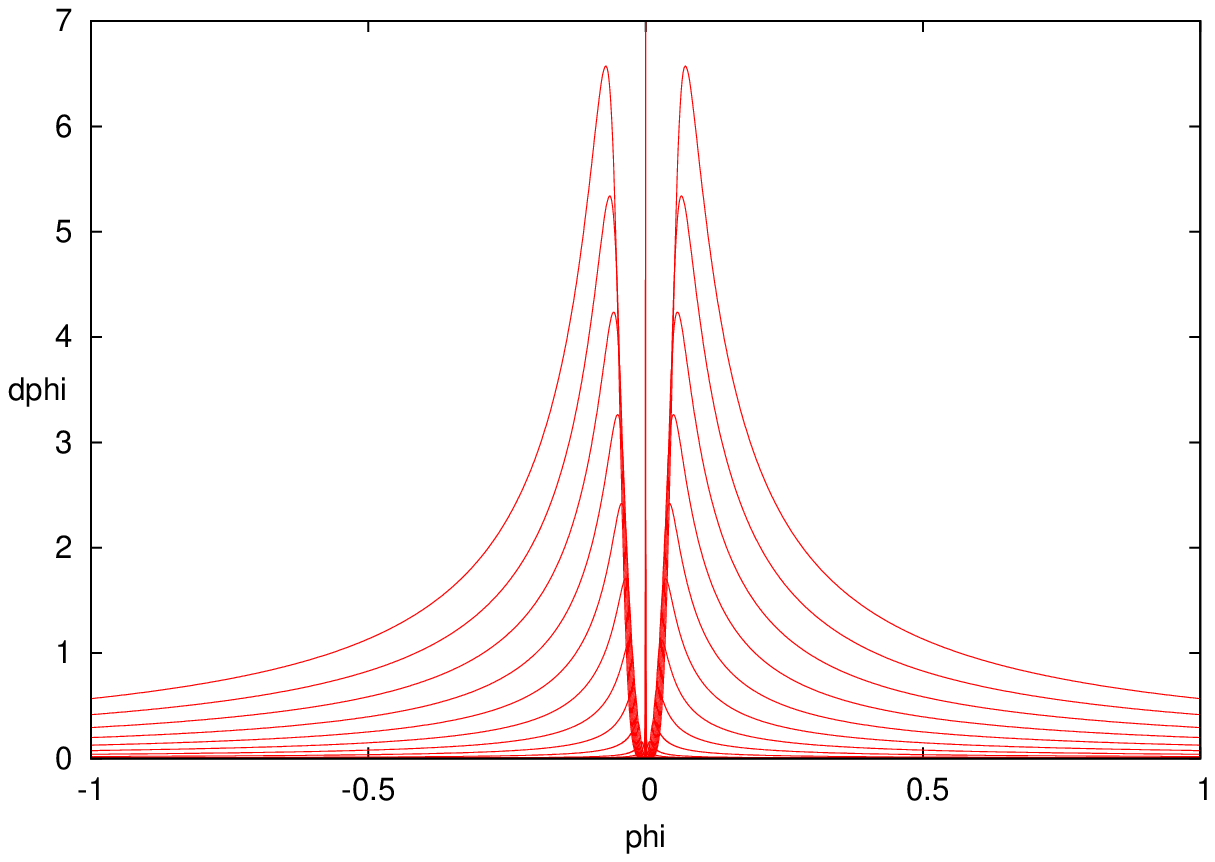}
    \label{fig:phase47K-zoom1}
  }
  \subfigure[]{
    \includegraphics[scale=0.60]{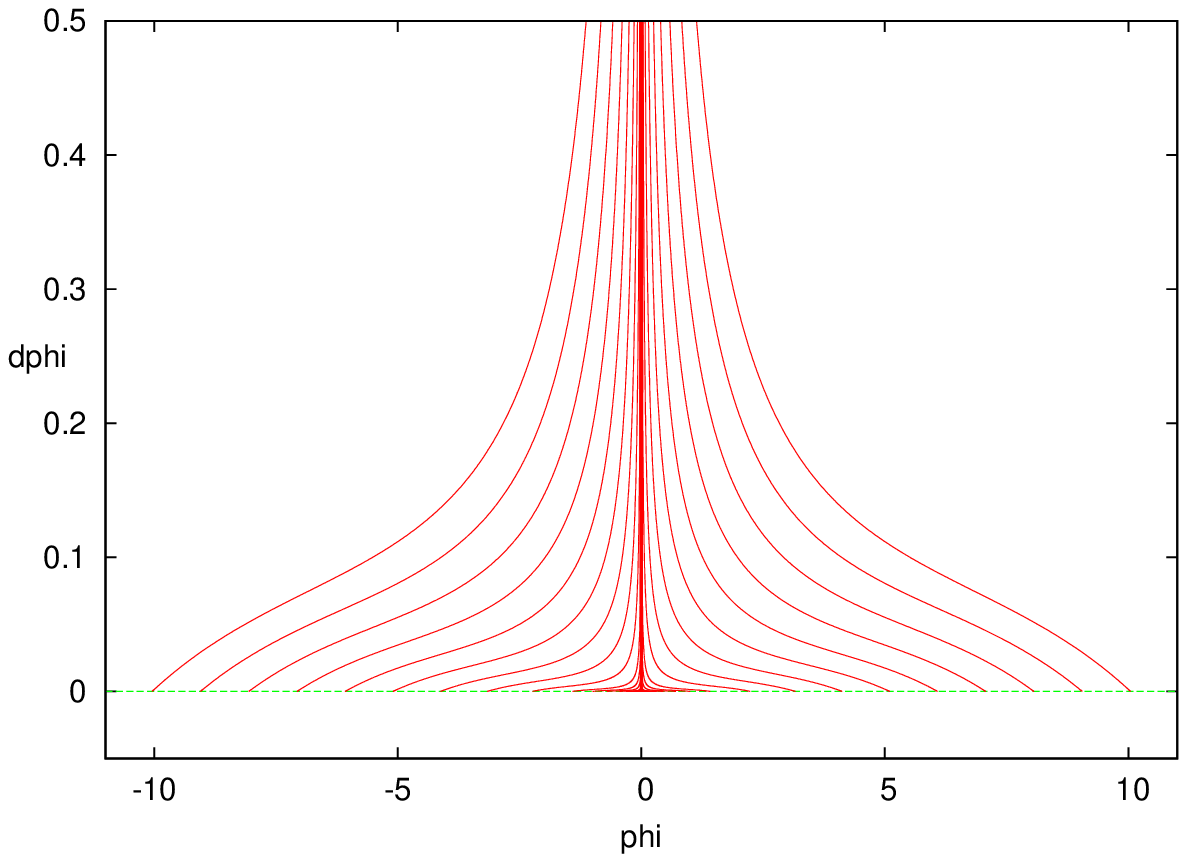}
    \label{fig:phase47K-zoom3}
  }
  \subfigure[]{
    \includegraphics[scale=0.60]{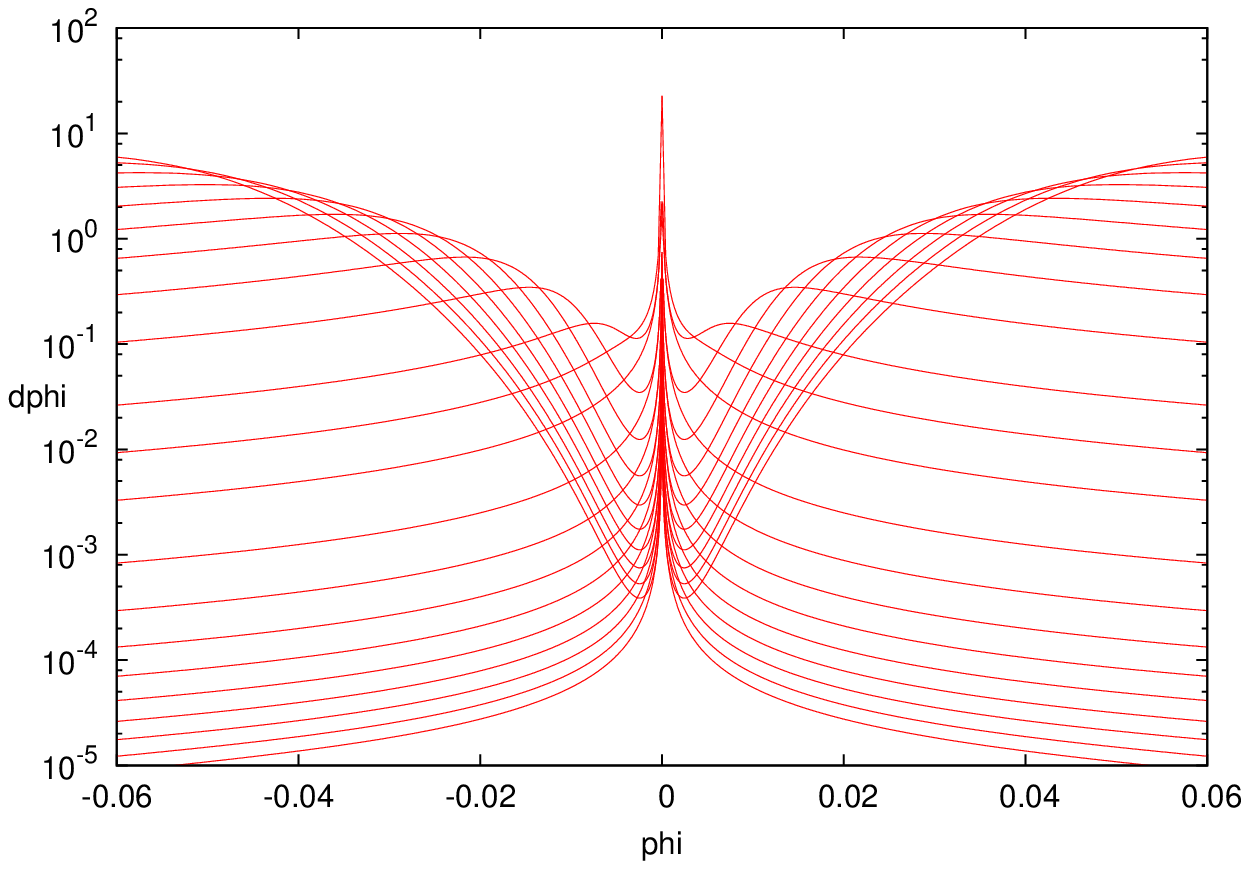}
    \label{fig:phase47K-zoom4}
  }
  \caption{Global view \ref{fig:phase47K-global-matt} and the zoom into particular 
    sectors of the phase portrait with respect to matter degrees of freedom for the 
    physical case of $\xi = 47000 \sqrt{\lambda}$. 
    In \ref{fig:phase47K-zoom1} we zoom the region around the bounce in the Einstein frame. 
    Figure \ref{fig:phase47K-zoom3} shows the trajectories at the onset of inflation 
    (at $\dot{\phi} = 0$), {providing the initial conditions for inflation}. 
    We do not show the attractor trajectory, as the length of inflation exceeeds the 
    technical (time) limits of our numerical simulations. 
    \ref{fig:phase47K-zoom4} shows the same sector as \ref{fig:phase47K-zoom1} in the 
    logarithmic scale to visualize the behavior of trajectories for small $\dot{\phi}$.
  }
  \label{fig:phase47K-globalandzooms}
\end{figure}

\section{Conclusions}\label{sec:concl}

% short summary of the work
{We investigated the model of non-minimally coupled scalar field coupled to 
gravity via additional term in the Lagrangian density: $\frac{1}{2}\xi\phi^2R$
and with potential $\frac{1}{4}\lambda\phi^4$. This system has been studied 
via effective dynamics methods within loop quantum cosmology framework. In the 
process of constructing the treatment we implemented the idea where the so-called 
Jordan frame represents the observed (physical) dynamics, however it is
an emergent framework coming from underlying one represented by Einstein frame.}
{The studies have shown that, similarly to other models within LQC, the high density 
region connects two semiclassical branches of the universe: contracting and 
expanding. However the process of the bounce itself is modified, featuring the 
``mexican hat'' shape of the scale evolution with highly non-adiabatic epoch 
of expansion$\to$contraction between two bounces.}
{The expanding epoch features the slow roll inflation whose length for all the 
investigated trajectories exceeds $60$ e-foldings.}
{These results, although solid within the specified model, cannot be treated as 
final due to {the model origin and its limitations. As we discussed in
sec.~\ref{sec:Einstein}, in the process of constructing the model we have made a {selection} 
of several alternatives in two aspects: which frame is the ``observed'' one 
and which is the correct starting point for quantization. The choices are not 
distinguished by any physical considerations. We simply tested one of few possible 
constructions in order to check whether it leads to unphysical results or, conversely, 
it is worth further investigation.}

% limitations of the work
{Let us now discuss the limitations:}
First, the quantization procedure applied here assumes the role of Jordan frame 
as the observed one and the Einstein frame as the underlying (somehow more 
fundamental) one. Such approach raises a (natural for non-minimally coupled 
systems) concern, that the description may admit a Lorentz symmetry violation, 
which in principle could imply disagreement with the cosmological observations
(for example, through the structure of the primordial gravitational waves). 
To the best of our knowledge there does not exist any definite result indicating 
that this would be the case but the issued in itself remains still open.

% here the trial and error discussion
{One also needs to remember that the above approach is just one of two possible in 
this case: quantization in Einstein and Jordan frame. Their quantization, especially 
the loop one may a priori lead to different dynamical results. At present there is 
no definite argument known which would select one of them as physical. In our studies 
we have taken the ``trial and error'' approach performing the ``physical viability'' test, 
i.e. checking whether the choice of the paricular frame would lead to immediate 
disagreements with known properties of the universe (and should thus be discarded) or, 
on the contrary, would give a promising point of departure for the model consistent 
with observations.} 
{In this paper we have selected for our studies Einstein frame, as 
for this case we have at our disposal the complete framework in context of full LQG. 
In consequence, the existing techniques of LQC could be applied directly.}
In the remaining case one needs to adapt the existing full LQG framework. This is a 
work currently in progress \cite{andrea-priv}. At present, it is already known 
that the modification of the formalism is essentially restricted to the fact of the 
constraints being of a different form (than the ``original'' LQG one), as expressed 
in terms of the adapted Ashtekar-Barbero variables. There is no indication of any 
conceptual problems preventing one from completing this framework. This expectation 
is further supported by the existing results in context of $f(R)$ and Brans-Dicke 
theories \cite{Zhang:2011gn,*Zhang:2011qq}. Of course, in context of the homogeneous 
LQC the necessary adaptation is much more straightforward. Early results of the LQC 
quantisation in Jordan frame suggest significant differences of the evolution of 
space-time around the bounce as well as different scale of the bounce in Jordan frame. 
Full analysis of the model considered here and comparison with the Einstein frame 
quantisation is the goal of the future work. 

{Furthermore, the {chosen} quantization procedure {in its exact form as specified 
in the paragraph above and in Sect.~\ref{sec:quantisation} is strictly tailored to 
the situations where the}}
gravitational part of the Hamiltonian {can} be expressed in the form where matter
is minimally coupled to gravity. However, {such} assumption cannot be satisfied
for a vector field non-minimally coupled to gravity (like models with vector inflation 
\cite{Golovnev:2008cf}), as the the transformation
described in the eq.~\eqref{eq:transformEin} would still leave non-minimal
coupling terms. {Nonetheless, one \emph{can} achieve the properties of the Einstein 
frame relevant for loop quantization of the geometry. Namely, the frame can be transformed 
to the one where the gravitational part of the action has the standard Einstein-Hilbert 
form. This is sufficient to apply standard loop techniques in their present form. 
In such case however the matter field would be still nonminimally coupled to gravity, 
which would require to use for the matter the quantization {schemes} of non-canonical 
kinetic terms followng from QFT (see for example \cite{Greenwood:2012aj}).}

{This problem (as well as the above solution to it) is present} even if a non-minimally 
coupled vector field is subdominant, i.e. when the presence of a vector field does 
not break isotropy of the background metric tensor. {The same applies to} any multi-inflationary 
scenarios with at least two fields non-minimally coupled to gravity.

% comments on the limits of effective dynamics
One has also to remember that the results presented in this paper have been 
obtained via the $0$th order effective dynamics. This choice of methodology 
is motivated on the one hand by the reliability of the method itself proven 
for other models of LQC and, on the other hand, by the aim of the study: 
verifying the robustness of sufficiently long inflation in the model considered
(which in turn required probing maximally wide region of the space of solutions).
However, the reported results can be treated as preliminary only, since 
the state-dependent parameters may affect the dynamics especially in the 
strongly nonadiabatic epoch between the bounces. Therefore, the effective method
has to be confirmed by a higher order effective treatment of \cite{bs-eff} 
and especially by the studies of the dynamics on the genuine quantum level 
within the framework of \cite{hp-lqc} 
(the latter in particular to determine the viable cutoff of the effective 
treatment order). These studies will be performed in future work.

Finally, let us note that the modification of the dynamics in the high energy 
epoch may also affect the structure of the inhomogeneities in the universe. 
{Within the treatment explored in this paper any classical perturbative 
framework should be constructed (and then subsequently quantized) in the Einstein 
frame generalized as discussed several paragraphs earlier. 
Then, (in particular)} the possible formation of shock waves in the strongly nonadiabatic 
epoch between bounces may in principle affect the structure formation in the 
expanding universe. Addressing these issues requires however extending the 
simple isotropic formalism used in this paper to inhomogeneous situations and further 
comparing the results with the analogus ones obtained in the Jordan frame directly.
The first step in {the latter} is the formulation of a well defined canonical formalism 
{(in particular the adapted Ashtekar-Barbero variables)} directly in the Jordan 
frame, followed next by the quantization 
within LQG framework. Then, an application of recent developments in the 
deparametrization techniques of LQG \cite{hp-lqg} will allow to unambiguously 
control the dynamical sector of the full theory.

This in turn will allow to construct a precise perturbative framework {(in both 
the generalization of the Einstein frame and the Jordan frame for comparizon)}. 
Methods of building such framework are already available in the context 
of LQC. They are: the \emph{hybrid quantization} \cite{mo-pert,*fmov-pert}, 
strictly implementing 
the unitary evolution, or \emph{Quantum Field Theory on Quantum Cosmological 
Spacetime} {\cite{akl-qft,*dlp-qft}}, already showing some preliminary successes in determining 
the primordial perturbations \cite{Nelson:2012ax,*Agullo:2012sh}. Applying 
these tools to the scenario studied here will ultimately allow to test its 
consistency with CMB observations and identify possible effects characteristic 
for it.

\begin{acknowledgments}
  We would like to thank Mikhail Shaposhnikov {and Dominika Konikowska} for 
  helpful discussions. The work was supported 
  in parts by grants of Minister Nauki i Szkolnictwa Wy{\.z}szego no.~182/N-QGG/2008/0, 
  N202~091839 and N202~104838 and by National Science Centre {of Poland} under research 
  grant DEC-2011/01/M/ST2/02466. The bulk of numerical simulations has been performed 
  with use of the \emph{Numerical LQC} library currently developed by T.~Paw{\l}owski, 
  {D.~Martin-deBlas} and J.~Olmedo.
\end{acknowledgments}

\appendix

\section{Deparametrization via massless scalar field}
\label{app:scalar-time}

One of the steps performed in order to arrive to an effective description of the 
considered system in Sec.~\ref{sec:Effective} was {the} introduction of the dust matter 
field and next the deparametrization of the system with respect to it. While this 
particular choice of the clock allows the full formal control over each step of 
{the construction of} the effective description, it is by no means {the only one possible}. 
Indeed, {deparametrization} can be performed (at the cost of {a limited} loss of precision) with respect 
to other matter fields {of} sufficiently good properties, for example the massless 
scalar considered in the original LQC works \cite{aps-imp} or the (appropriately 
tailored) {population of Maxwell fields} \cite{ppwe-rad}. As long as we restrict the expansion 
to the $0$th order and assume vanishing mean energy of the clock field, the formalism 
leads to exactly the same equation of motion \eqref{eq:eoms-Einst} as in the case 
of dust clock.
In this appendix we briefly outline the construction of the effective description 
with use of the massless scalar field. {The exact same procedure can be applied 
to Maxwell fields of \cite{ppwe-rad}.}

Our starting point is the modification of the classical Hamiltonian constraint 
in the Einstein frame, {namely} adding to it the massless scalar field term:
\begin{equation}
  {\bf H} \to {\bf H}' = {\bf H} + {\bf H}_{\varphi} \ , \qquad
  {\bf H}_{\varphi} = \frac{p_{\varphi}^2}{2\alpha |v|} \ ,
\end{equation}
where $p_{\varphi}$ is the canonical momentum of the clock field $\varphi$ and 
the variable $v$ (and the constant $\alpha$) is defined {via eq.~\eqref{eq:V-def}}.
This constraint is next deparametrized with respect to $\varphi$\footnote{See 
  \cite{aps-imp} of the detailed description in the case {of vacuum with 
  massless scalar field.}
}, {in effect} being brought to the form
\begin{equation}
  p_{\varphi}^2 = \Theta = -2\alpha |v|{\bf H} \ ,
\end{equation}
which in turn can be understood of as {representing} the free system evolving 
with respect to $\varphi$. Applying polymer quantization {as in}
Sec.~\ref{sec:quantisation} (see \cite{aps-imp} {for further information}) 
we arrive to the generalized Klein-Gordon equation
\begin{equation}
  - \partial_{\varphi}^2\Psi(v,\phi) = \hat{\Theta}\Psi(v,\phi) \ , \qquad
  \hat{\Theta} = -2\alpha :\!|\hat{v}|\hat{{\bf H}}\!: \ , 
\end{equation}
where $:\cdot:$ means the symmetric ordering. Action of $\hat{\Theta}$ on a
dense domain in $\Hil_{\rm kin}$ reads (following \cite{acs-slqc} for its
gravitational part)
\begin{subequations}\begin{align}
  \hat{\Theta} &= \hat{\Theta}_{\rm gr} \otimes \id_{\phi} - \hat{\Theta}_{\phi}
       \ ,\\
  \hat{\Theta}_{\rm gr} 
    &= -\frac{3\pi G}{4}
  \left[\sqrt{|\hat{v}|}(\hat{N}-\hat{N}^{-1})\sqrt{|\hat{v}|}\right]^2 \ , \qquad 
  \hat{\Theta}_{\phi} = \left[ \id_{\rm gr}\otimes \hat{\pi}_{\tilde{\phi}}^2 
    + 2\alpha^2\hat{v}^2\otimes\tilde{V}(\tilde{\phi}) \right] \ .
\end{align}\end{subequations}
Taking one of the superselected sectors {(e.g. the one corresponding to the positive
frequencies)}, one arrives to the \emph{formal} {Schr\"odinger} equation
\begin{equation}
  i \partial_{\varphi}\Psi(v,\phi) = \sqrt{\hat{\Theta}}\Psi(v,\phi) \ .
\end{equation}
To write it precisely one needs to overcome several obstacles as $\hat{\Theta}$ 
is neither positive definite nor essentially self-adjoint \cite{kp-posL}, since 
it contains the potential term acting as effective positive cosmological
constant on $\phi=\const$ plane. It is however expected to admit a family of 
non-unique selfadjoint extensions. Therefore one has to choose one extension 
$\hat{\Theta}_{\beta}$ (where $\beta$ is an abstract extension label). Then, 
upon fixing the gauge $t=\varphi$, the Schroedinger equation can be written as
\begin{equation}
  i \partial_{t}\Psi(v,\phi) = \hat{{\bf H}}_{\beta}\Psi(v,\phi) 
  := \sqrt{[\hat{\Theta}_{\beta}]_+}\Psi(v,\phi) \ ,
\end{equation}
where $[\cdot]_+$ denotes the positive part of the operator.
Since only the positive part of $\hat{\Theta}_{\beta}$ enters the Hamiltonian, 
the physical Hilbert space does not coincide with $\Hil_{\kin}$. Instead, it is 
its proper subspace $\Hil_\beta$ spanned by the eigenfunctions of $[\hat{\Theta}_{\beta}]_+$. 
In principle, {different extensions could produce different physical subspaces}, 
as it is the case for positive cosmological constant \cite{ap-posL}. 
The consequence of the latter is that every operator defined on $\Hil_{\kin}$
has to be further projected on $\Hil_{\beta}$ to define the physical operator,
thus the operators like $\hat{v}$ lose their simple analytical form.

Proceeding with the construction of the effective description requires a heuristic
assumption: namely, that for sufficiently large class of states the projection
does not affect the action of the elementary operators significantly. Given
that, we can replace $\hat{v}^n$ and $\hat{N}$ by their expectation values,
arriving to the Hamiltonian
\begin{equation}\label{eq:Heff-phi}
  {\bf H}_{\eff} = \sqrt{3\pi G v^2\sin^2(b) - \pi_{\tilde{\phi}}^2 
    - 2\alpha^2 v^2 \tilde{V}(\tilde{\phi})} = p_{\varphi} =: E_{\varphi} \ .
\end{equation}
Similarly to the dust time case, this (also heuristic) step corresponds to the
$0$th order effective expansion introduced in \cite{bs-eff}. Here, however, this 
procedure is not fully justifiable, since it ignores the effects of inequivalent 
self-adjoint extensions on the dynamics (see the discussion in \cite{bbhkm-eff}) 
-- effects that, although small, are usually present \cite{ap-posL}. 

{If we accept} the heuristic steps discussed above, we arrive to an effective model
described by the Hamiltonian \eqref{eq:Heff-phi}. To compare it against the
treatment of Sec.~\ref{sec:Effective} we note that:
\begin{enumerate}[(i)]
  \item For $p_{\varphi}\neq 0$ the model is equivalent to the one described \
    by the Hamiltonian 
    \begin{equation}\label{eq:Heff-phi2} 
      {\bf H}_{\eff}' = 3\pi G v^2\sin^2(b) - \pi_{\tilde{\phi}}^2 
	- 2\alpha^2 v^2 \tilde{V}(\tilde{\phi}) \ .
    \end{equation}
    The equations of motion of this model are equivalent to the one of
    \eqref{eq:Heff-phi}
    with the lapse function rescaled via $N \to 2p_{\varphi}N$.
  \item Under the change of sign and the lapse function transformation 
    $N\to N/(2\alpha v)$ the
    Hamiltonian \eqref{eq:Heff-phi2} transforms into the effective Hamiltonian 
    of the system with dust time \eqref{eq:Heff}.
\end{enumerate}
In consequence, at the level of $0$th order effective dynamics the deparametrization 
with respect to massless scalar field and irrotational dust are physically equivalent.

\bibliography{adp-nonminimal-arXiv-v2}{}
\bibliographystyle{apsrev4-1}

\end{document}